\documentclass[%
reprint, 
 superscriptaddress,
showpacs,preprintnumbers,
 amsmath,amssymb,
 aps,
 prd,
]{revtex4-1}

\usepackage{graphicx}
\usepackage{dcolumn}
\usepackage{bm}
\usepackage{color}

\usepackage{cancel}
\usepackage{multirow}
\newcommand{\met}{\cancel{E}_T}

\begin{document}

\preprint{DESY 17-035}

\title{Signatures of Dirac and Majorana Sterile Neutrinos in Trilepton Events at the LHC}

\author{Claudio O. Dib}
\email{claudio.dib@usm.cl}
\affiliation{ CCTVal and Department of Physics, Universidad T\'ecnica Federico Santa Mar{\'{\i}}a, Valpara{\'{\i}}so, Chile }

\author{C. S. Kim}%
\email{cskim@yonsei.ac.kr}
\affiliation{ Department of Physics and IPAP, Yonsei University, Seoul 120-749, Korea }%

\author{Kechen Wang}
\email{kechen.wang@desy.de (corresponding author)}
\affiliation{ DESY, Notkestraße 85, D-22607 Hamburg, Germany }%
\affiliation{ Center for Future High Energy Physics, Institute of High Energy Physics, Chinese Academy of Sciences, Beijing, 100049, China }%


\begin{abstract}
Heavy sterile neutrinos with masses below $M_W$ can induce trilepton events at the 14 TeV LHC through purely leptonic $W$ decays of $W^\pm \to e^\pm  e^\pm  \mu^\mp \nu$ and $\mu^\pm  \mu^\pm  e^\mp \nu$ where the heavy neutrino will be  in an intermediate state on its mass shell. Discovery and exclusion limits for the heavy neutrinos are found using both Cut-and-Count (CC) and a Multi-Variate Analysis (MVA) methods in this study.
We also show that it is possible to discriminate between a Dirac and a Majorana heavy neutrino, even when lepton number conservation cannot be directly tested due to unobservability of the final state neutrino. This discrimination is done by exploiting a combined set of kinematic observables that differ between the Majorana vs. Dirac cases. We find that the MVA method can greatly enhance the discovering and discrimination limits in comparison with the CC method.
For a 14-TeV $pp$ collider with integrated luminosity of 3000 ${\rm fb}^{-1}$, sterile neutrinos can be found with 5$\sigma$ significance if heavy-to-light neutrino mixings $|U_{Ne}|^2 \sim |U_{N\mu}|^2 \sim 10^{-6}$, while the Majorana vs. Dirac type can be distinguished if $|U_{Ne}|^2 \sim |U_{N\mu}|^2 \sim 10^{-5}$ or even $|U_{N\ell}|^2\sim 10^{-6}$ if one of the mixing elements is at least an order of magnitude smaller than the other.
\end{abstract}

\maketitle


\section{Introduction}
\label{sec:intro}

Neutrinos are the most esoteric of all  particles in the Standard
Model (SM):
they interact via weak interactions only, which makes them very hard to detect;
their interaction is purely of left handed chirality, so their right handed component --if there is any-- is sterile;
their masses, while not all zero, are much smaller than the energies of all detectable processes, so they are only available as extremely relativistic particles; they exist in three flavors, pairing the charged leptons, however they exhibit large mixing in their mass eigenstates, thus exhibiting a chameleonic
behaviour by changing their flavor in flight, which is known as neutrino
oscillations.
The very observation of neutrino oscillations~\cite{Oscillation_experiment} implies that neutrinos must have 
mass, contrary to the Standard Model (SM) in which the neutrinos are assumed to be massless.
Moreover, since neutrinos are electrically neutral, they could be their own antiparticles, i.e.  Majorana fermions~\cite{Majorana_theory}, carrying no charge such as lepton number, in which case weak interactions involving neutrinos will not conserve lepton number. Alternatively, if they carry lepton number, they must be Dirac fermions, neutrinos and antineutrinos will be different particles, and lepton number will be conserved.  
Therefore, one important step towards resolving the origin of neutrino mass is to ascertain
whether they are Dirac or Majorana fermions. The Majorana nature of neutrinos can be revealed in neutrino-less double beta  decay ($0\nu\beta\beta$)
experiments, but so far no evidence has been found~\cite{Majorana_experiments, Engel:2016xgb}.
Concerning explanations of the smallness of neutrino masses, most are based on 
seesaw mechanisms~\cite{Minkowski:1977sc, Ramond:1979py, Yanagida:1979as, Mohapatra:1979ia, Glashow, Schechter:1980gr, Magg:1980ut, Cheng:1980qt, Lazarides:1980nt, Foot:1988aq}, 
which imply the existence of additional, heavier neutrinos, which are sterile under the electroweak interactions except for their mixings with the standard neutrinos~\cite{Other_proposals}. 
%
The original seesaw models required very large masses for the sterile neutrinos, $M\sim 10^5$ to $10^{15}$ GeV, beyond detectability in any foreseen experiment, and  mixings  $U\sim (m_\nu/M)^{1/2}$ which are highly suppressed as well  ($10^{-8}$ to $10^{-13}$). However, in other versions called \emph{low-scale seesaw, inverse seesaw, etc.} the smallness of $m_\nu$ does not require huge values for $M$ nor tiny values for $U$. To date, each specific scenario proposes heavy neutrinos with their mass within a given scale, but from one scenario to another this scale can be anywhere from a few eV all the way to grand unification scales. In turn, different experiments put bounds on neutrino masses and mixings, each one in a different and limited mass range within this broad spectrum of possibilities. 
%
%
So far, experimental searches have not found conclusive
evidence of their existence~\cite{Other_experiments} neither as Dirac or Majorana particles.
In particular the studies at the Large Hadron Collider (LHC) look for same sign dilepton plus dijet events, $\ell^\pm\ell^\pm j j$, which can be produced and observed 
if there are heavy Majorana fermions with mass above a few tens of GeV 
and up to a few hundred GeV~\cite{cmsmaj, atlasmaj}. 
For neutrino masses below $M_W$, the jets may not be energetic enough to be separated from the background and thus trilepton events $\ell^\pm\ell^\pm\ell^{\prime\mp}\nu$ would provide clearer signals~\cite{Izaguirre:2015pga}. 

In previous works we have studied the potential of these trilepton events to discover heavy neutrinos, especially 
addressing the discrimination
between their Dirac or Majorana nature~\cite{Dib:2015oka}.  
We studied the signal $W^\pm \to e^\pm e^\pm \mu^\mp \nu$, which will appear resonantly enhanced provided there exist neutrinos with masses below $M_W$, through the subprocess $W^\pm \to e^\pm N$ followed by
$N\to e^\pm\mu^\mp\nu$ (where $N$ stands for the heavy neutrino). The choice of having no opposite-sign same-flavor (no-OSSF) lepton pairs in the final state helps eliminate a serious SM radiative background  $\gamma^*/Z \to \ell^+\ell^-$ \cite{p2mu}.
If $N$  is Majorana, it will induce a lepton number conserving (LNC) process $W^+ \to e^+ e^+ \mu^- \nu_e$  as well as a lepton number violating (LNV) process $W^+ \to e^+ e^+ \mu^- \bar\nu_\mu$, while if it is of Dirac type, it will induce the LNC process only. One could in principle use this feature to discriminate between a Majorana and a Dirac $N$, 
however, since the final neutrino escapes detection, the observed final state is just $e^\pm e^\pm \mu^\mp$ or $\mu^\pm \mu^\pm e^\mp$ { plus missing energy. It is then not a simple task} to distinguish between the LNC and the LNV processes, and hence between the Majorana vs. Dirac nature of $N$.  In our previous work we found that, in principle, the two cases could be distinguished by constructing the energy spectrum of the opposite-charge lepton.

In our consecutive work \cite{Dib:2016wge}, we presented a simpler method 
to distinguish between Majorana and Dirac $N$'s,
by examining the full decay rates instead of the spectra, for all the channels $e^\pm e^\pm \mu^\mp$ and $\mu^\pm \mu^\pm e ^\mp$, because the discrimination through spectra in rare processes is much more difficult to achieve. However, 
this discrimination, which is based on full rates, only works if  the mixing parameters $U_{Ne}$ and $U_{N\mu}$ are considerably different from each other.

Here we present a comprehensive strategy to detect heavy sterile neutrinos and discriminate between Dirac vs. Majorana using all the details of the leptonic decays of $W^\pm \to e^\pm  e^\pm  \mu^\mp \nu$ or $\mu^\pm  \mu^\pm  e^\mp \nu$ at the 14 TeV LHC,
provided they exist with masses near and below the $W$ boson mass. The strategy is based on both a Cut-and-Count (CC) method and a Multi-Variate Analysis (MVA) method that uses all the features of the leptonic decays that can distinguish between Dirac and Majorana neutrinos.

The article is organized as follows. In Section~\ref{sec:theoCal} we present the theoretical aspects and formulation of the problem.
In Section~\ref{sec:simuBG} we describe the data simulation and background study. In Section~\ref{sec:NvsBG} we state the method to detect the existence of heavy neutrinos that induce the trilepton events. In Section~\ref{sec:drcVSmaj} we study the capability of the method to discriminate between the Dirac or Majorana character of $N$. In Section~\ref{sec:Summary} we present our summary and conclusions.


\section{Theoretical framework}
\label{sec:theoCal}

\begin{figure}[h]
\includegraphics[scale=0.7]{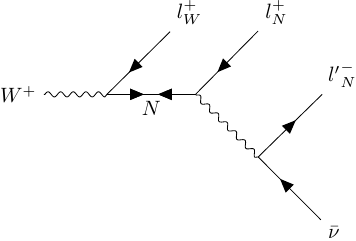}
\includegraphics[scale=0.7]{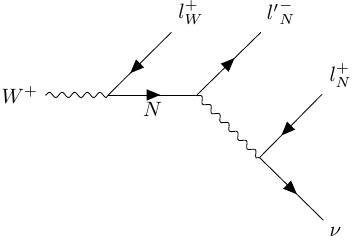}
\caption{ Left: LNV process $W^+ \to l_W^+ l_N^+ {l^\prime}_N^- \bar\nu$,  mediated by a heavy sterile neutrino of Majorana type only. Right: 
 the LNC process $W^+\to l_W^+ {l^\prime}_N^- l_N^+ \nu$, mediated by a heavy sterile neutrino of either Majorana or Dirac type.}
\label{fig:Wdecay}
\end{figure}

Our processes of interest are labelled as $W^{\pm} \to l_{W}^{\pm} l_{N}^{\pm} {l^\prime}_{N}^{\mp} \nu$ and depicted in Fig.~\ref{fig:Wdecay}. Here $l$ and $l^\prime$ are charged leptons of different flavours, either $e$ or $\mu$ (i.e. $e^\pm  e^\pm  \mu^\mp \nu$ or $\mu^\pm  \mu^\pm  e^\mp \nu$), where $\nu$ represents a SM neutrino or anti-neutrino.
These are two different processes, but since the final neutrino goes undetected, the observable final state is the same, namely trilepton $l^{\pm} l^{\pm} {l^\prime}^{\mp}$ {plus missing energy}, in either process. One process is lepton number violating (LNV), while the other is lepton number conserving (LNC). 
If $N$ is Majorana, both LNV and LNC processes occur and the events of the two modes add up in the experiment, while if $N$ is Dirac, only the LNC process occurs. 

Following the notation of Fig.~\ref{fig:Wdecay} for the lepton momenta and calling $q$ the momentum of the $W$ boson, the differential rate of the LNV process with flavors {\it e.g.} $e^+e^+\mu^-$, is:
 \begin{eqnarray}
&&\Gamma (W^+\to e^+ e^+\mu^-\bar\nu_\mu) =
\frac{8\sqrt{2} G_F^3 }{3\pi}  \frac{m_N(M_W^2-m_N^2)}{M_W\Gamma_N}
|U_{N e}|^4 
\nonumber\\  
&& 
\int d\Phi_3 \  (\ell_N \cdot \ell_\nu)  \Big\{ ( \ell_W \cdot \ell^\prime_N ) + \frac{2}{M_W^2}  (q\cdot \ell_W) 
(q\cdot \ell^\prime_N )  
   \Big\} ,
   \label{LNVdist}
\end{eqnarray}
where $\Gamma_N$ is the $N$ width, which also depends on $m_N$ and the lepton mixings 
\cite{Dib:2015oka}. We denote by  
$\int d\Phi_3$ the Lorentz invariant phase space for the three final particles of the $N$ decay in the normalization of the Particle Data Group \cite{PDG}.
Similarly, the differential rate for the LNC process is: 
\begin{eqnarray}
&&\Gamma (W^+\to e^+ e^+\mu^-\nu_e) = 
\frac{8\sqrt{2} G_F^3 }  {3\pi } \frac{(M_W^2 - m_N^2)}{m_N M_W \Gamma_N}
|U_{N e}U_{N \mu}|^2 
\nonumber\\
&&   
\times \int d\Phi_3  \ \  (\ell^\prime_N  \cdot \ell_\nu)   \times  \Bigg\{ 
2 (k_N\cdot \ell_N) \Big[ ( k_N \cdot \ell_W )
\nonumber \\ 
&&
\hbox{\hspace{3.5cm}}
 + \frac{2}{M_W^2}  (q\cdot k_N )  (q\cdot \ell_W ) \Big] 
\nonumber \\ 
&&
\hbox{\hspace{1cm}}
- m_N^2 \Big[   (\ell_W \cdot \ell_N) + \frac{2}{M_W^2}  (q\cdot \ell_W )  (q\cdot \ell_N) \Big]
   \Bigg\} .
   \label{LNCdist}
\end{eqnarray}
These two expressions seem to be different. Indeed, it is straightforward to show that the two processes have different spectral and angular distributions. Their integrated branching ratios, on the other hand, are equal except for a global lepton mixing factor \cite{Dib:2015oka}:

 \begin{eqnarray}
 {\rm Br}(W^+\to e^+ e^+\mu^-\bar\nu_\mu) &=& f(m_N)  \times \frac{|U_{Ne}|^4}{\sum_{\ell} |U_{N\ell}|^2},
 \label{LNVrate}
 \\
 {\rm Br}(W^+\to e^+ e^+\mu^-\nu_e) &=& f(m_N) \times \frac{|U_{Ne}|^2|U_{N\mu}|^2}{\sum_{\ell} |U_{N\ell}|^2},
 \label{LNCrate}
 \end{eqnarray}
where
\[
f(m_N) \approx
 4.8 \times 10^{-3} \   
 \left(1 - \frac{m_N^2}{M_W^2}\right)^2
 {\left( 1 + \frac{m_N^2}{2M_W^2}\right) }  .
\]

The spectral distributions were studied in a previous work \cite{Dib:2015oka}. 
Concerning the angular distributions, from Eqs. (\ref{LNVdist}) and (\ref{LNCdist}), one can see that the LNV and LNC processes differ in their combinations of scalar products, which translates into different angular distributions among pairs of particles. Previous works have exploited these angular distribution differences for other models \cite{Han:2012vk}. In our case, each one of these differences are not dramatic by themselves, but in our analysis we build a combination of several distributions that differ between the LNV and LNC modes, adding up in their capacity to discriminate between a Dirac and a Majorana $N$. 

For convenience, we introduce two parameters: the normalization factor ``$s$'' and the disparity factor ``$r$'':
\begin{equation}
s \equiv 2\times10^6\, \frac{|U_{Ne} U_{N\mu} |^2}{|U_{Ne}|^2+|U_{N\mu}|^2} ,\,\,\,\,\, r \equiv \frac{|U_{Ne}|^2}{|U_{N\mu}|^2} .
\end{equation}
The mixing angles $|U_{Ne}|^2$ and $|U_{N\mu}|^2$ can be expressed in terms of $r$ and $s$ as
\begin{equation}
|U_{Ne}|^2 = \frac{s\, (1 + r)}{2 \times 10^6},\,\,\,\,\, |U_{N\mu}|^2 = \frac{s\, (1 + \frac{1}{r} )}{2 \times 10^6}.
\end{equation}

In our study we assume for simplicity that only one sterile neutrino $N$ is within the experimental reach, and that it mixes with the active neutrinos $\nu_e$ and $\nu_{\mu}$ only. The sterile neutrino can be either Dirac or Majorana.
According to Eqs. (\ref{LNVrate}) and (\ref{LNCrate}), the branching ratios of $W$ decaying to trilepton final states via the sterile neutrino in the LNV processes go as:
\begin{eqnarray}
{\rm Br}(W^{\pm} \to e^{\pm} e^{\pm} \mu^{\mp} \nu) &\propto &  
s\times r \, , \nonumber \\
{\rm Br}(W^{\pm} \to \mu^{\pm} \mu^{\pm} e^{\mp} \nu) &\propto & 
\frac{s}{r} \, .
\end{eqnarray}
while the braching ratios in the LNC processes go as:
\begin{eqnarray}
{\rm Br}(W^{\pm} \to e^{\pm} e^{\pm} \mu^{\mp} \nu) &\propto & 
s \, , \nonumber \\
{\rm Br}(W^{\pm} \to \mu^{\pm} \mu^{\pm} e^{\mp} \nu) &\propto & 
s \, ,
\end{eqnarray}
Therefore, the production rates for the Dirac case (LNC process only) and Majorana case (both LNC and LNV processes) corresponding to different trilepton final states are proportional to the scale factors shown in Table~\ref{tab:Nsig}.

\begin{table}[h]
\centering
\begin{tabular}{ccc}
\hline
\hline
  & Dirac & Majorana   \\
\hline
$e^{\pm} e^{\pm} \mu^{\mp} \nu$ & $s$ & $s\,(1+r)$ \\
$\mu^{\pm} \mu^{\pm} e^{\mp} \nu$ & $s$ & $s\,(1+ 1/r )$ \\
\hline
\hline
\end{tabular}
\caption{Scale factors due to lepton mixing, for the production rates of the different trilepton modes.}
\label{tab:Nsig}
\end{table}

In the following section we present our studies of simulated events in $pp$ collisions at 14 TeV at the LHC within this theoretical framework, including SM backgrounds.


\section{Data simulations and background studies}
\label{sec:simuBG}

In this section, we describe in detail our event simulations, the observables which can be used to reject the SM backgrounds, and our strategies to determine the discovery potential of a heavy sterile neutrino $N$ with mass below $M_W$ and to determine its Majorana or Dirac character.
For the data simulation, similar to our previous work \cite{Dib:2016wge}, 
we build a Universal FeynRules  Output \cite{Degrande:2011ua} model file that extends the SM with additional sterile neutrino interactions using $\texttt{FeynRules}$ \cite{Christensen:2008py} and implement it into $\texttt{MadGraph 5}$ \cite{Alwall:2014hca} to generate the signal events. We explore two benchmark points: $m_N = $ 20 GeV and 50 GeV, both with $r = s = 1$ (i.e., $|U_{Ne}|^2 = |U_{N\mu}|^2 = 10^{-6}$). The background events are also 
generated with $\texttt{MadGraph 5}$.
The parton showering and hadronization are finished with $\texttt{PYTHIA 6}$ \cite{Sjostrand:2006za}, while the detector simulation is completed with the help of $\texttt{DELPHES 3}$ \cite{deFavereau:2013fsa}. At the parton level, we include up to two extra partons for both signal and background processes, and the jet matching is performed using the MLM-based shower-$k_\bot^{}$ scheme (named after Michelangelo L. Mangano --see Ref. \cite{Alwall:2008qv}).
To maintain consistency through all our study, the production cross sections calculated by $\texttt{MadGraph 5}$ are used to estimate the number of events for both signal and background processes as well.

Although in this trilepton search we demand no lepton pairs with opposite sign and same flavor in the final state (no-OSSF)  in order to reject backgrounds from radiative pairs, there still exists non-negligible backgrounds from various processes.
The dominant SM backgrounds can be divided into two categories: (i) from leptonic $\tau$ decays and (ii) from fake leptons.
In the first category, the dominant process is the pair production of $WZ$ with $W$ decaying leptonically and $Z \to \tau\tau$. The trilepton final states with no-OSSF pairs can arise from the subsequent leptonic decay of $\tau$'s. We estimate this background process via Monte Carlo (MC) simulations.

The dominant processes of the second category are $\gamma^*/Z$+jets and $t\bar{t}$, where two leptons come 
from $\gamma^*/Z \to \tau\tau$ or the prompt decay of $t$ and $\bar{t}$, and a third lepton is faked from jets containing heavy-flavor mesons. 
Although in general {fake leptons} from such heavy-flavor meson decays are not well isolated, there are still rare occasions when they can pass the lepton isolation criteria \cite{Sullivan:2006hb,Sullivan:2008ki,Sullivan:2010jk}.  
Because these background processes ($\gamma^*/Z$+jets and $t\bar{t}$) have large cross sections and small fake probabilities, it is very challenging to obtain enough statistics for background study in a purely MC simulation. Moreover, simulating such processes requires a detailed modeling of the jet fragmentations, and current level of MC simulation may not be accurate enough. For these reasons, data-driven methods are used by the ATLAS and CMS collaborations to estimate the fake lepton contributions \cite{Khachatryan:2015gha,Chatrchyan:2014aea,ATLAS:2014kca}, a matter beyond the scope of this study.

In this work, similar to our previous study~\cite{Dib:2016wge}, we adopt a phenomenological \emph{fake lepton} (FL) simulation method, originally introduced in Ref.~\cite{Curtin:2013zua} and later also implemented in
Ref.~\cite{Izaguirre:2015pga}. 
A fake lepton originates from  a jet that generates an imprint in the detector that resembles that of a lepton, and therefore it inherits part of the kinematics of the actual jet. For the FL simulation, two modeling functions are introduced: a \emph{mistag efficiency}, $\epsilon_{j \rightarrow \ell}^{}$, which is the probability of a particular jet to be faked as a lepton, and a  
\emph{transfer function}, $\mathcal{T}_{j\rightarrow \ell}^{}$, which is a probability distribution that determines how much of the jet momentum is transferred into the fake lepton. These two functions contain some modeling parameters that can be fitted by validating simulated results against those of the actual experiment. We revisit the validation performed in Ref.~\cite{Izaguirre:2015pga}, and found that the modeling parameters they obtained are consistent with the experimental results. Thus, the same set of parameters are used here. We also assume the same fake efficiency for electrons and muons. Details of this FL simulation method and the validation can be found in Refs.~\cite{Izaguirre:2015pga,Curtin:2013zua}. Our validation results are shown in Appendix \ref{sec:FL}.

For the analysis, we first select the events with 3 leptons $l^{\pm} l^{\pm} {l^\prime}^{\mp}$ which have no lepton pairs with Opposite Sign and Same Flavor (no-OSSF). Then, the following basic cuts for the leptons and jets are applied: $p_{T,l} \geq$ 10 GeV and $|\eta_l| \leq$ 2.5; $p_{T,j} \geq$ 20 GeV and $|\eta_j| \leq$ 5.0. We also veto b-jets to suppress the $t\bar{t}$ background.

Then, to pick up the correct lepton from the $N$ decay in the Same Sign Same Flavor (SSSF) lepton pair ($l^{\pm} l^{\pm}$) for the rest of the analysis, we construct a chi-square function
\begin{equation}
\chi^2= \left( \frac{M_W - m_W}{\sigma_W} \right) ^2 + \left( \frac{M_N - m_N}{\sigma_N} \right)^2,
\label{eqn:Chi2}
\end{equation}
where $m_W$ is the input $W$ mass of 80.5 GeV; $m_N$ is the assumed $N$ mass (20 or 50 GeV in our benchmarks), while $M_W$ and $M_N$ are the reconstructed $W$ and $N$ masses from the invariant mass of the $l^{\pm} l^{\pm} {l^\prime}^{\mp} \nu$ and $l^{\pm} {l^\prime}^{\mp} \nu$ systems, respectively; $\sigma_W$ and $\sigma_N$ denote the widths of the reconstructed mass distributions, which 
are taken
to be 5\% of the respective  $m_W$ and $m_N$, for simplicity. When calculating the reconstructed mass $M_W$ and $M_N$, the final neutrino transverse momentum ${\overrightarrow{p}}_{T,\nu}$ is assumed to be the missing transverse momentum. The neutrino longitudinal momentum $p_{z,\nu}$ and the correct lepton $l^{\pm}$ from the $N$ decay are determined by minimizing the $\chi^2$ of Eq.~(\ref{eqn:Chi2}).

\begin{figure}[h]
\includegraphics[scale=0.09]{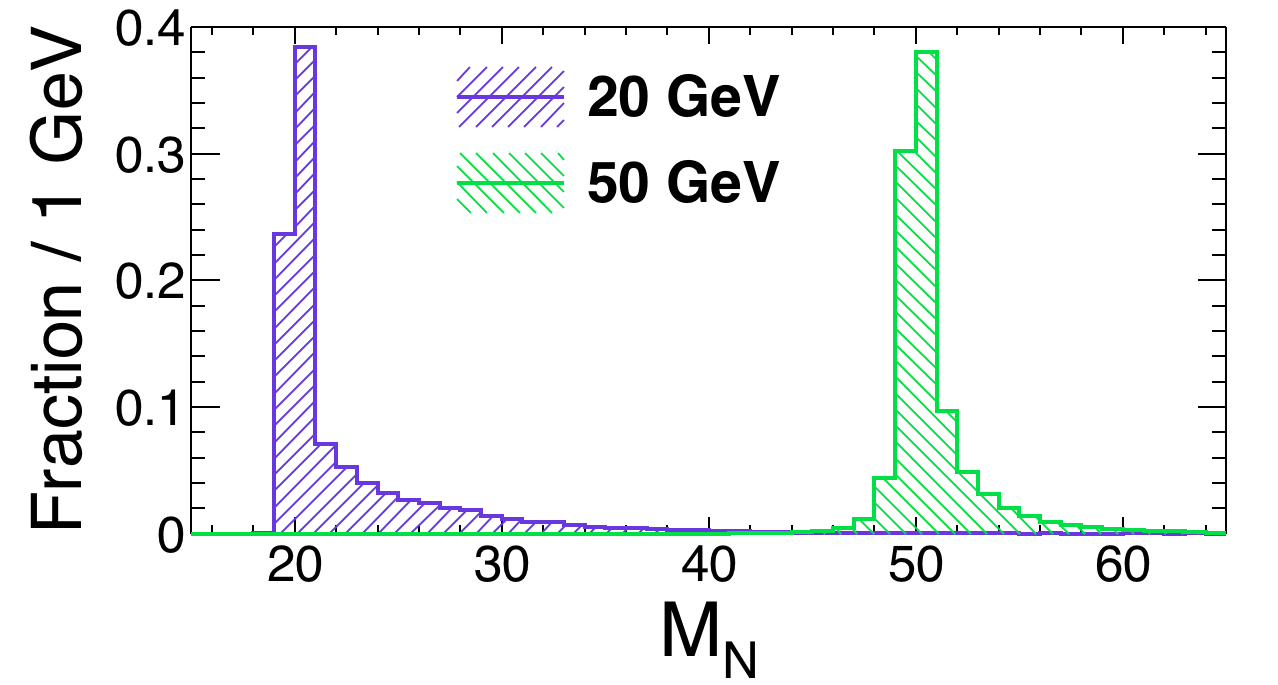}
\includegraphics[scale=0.09]{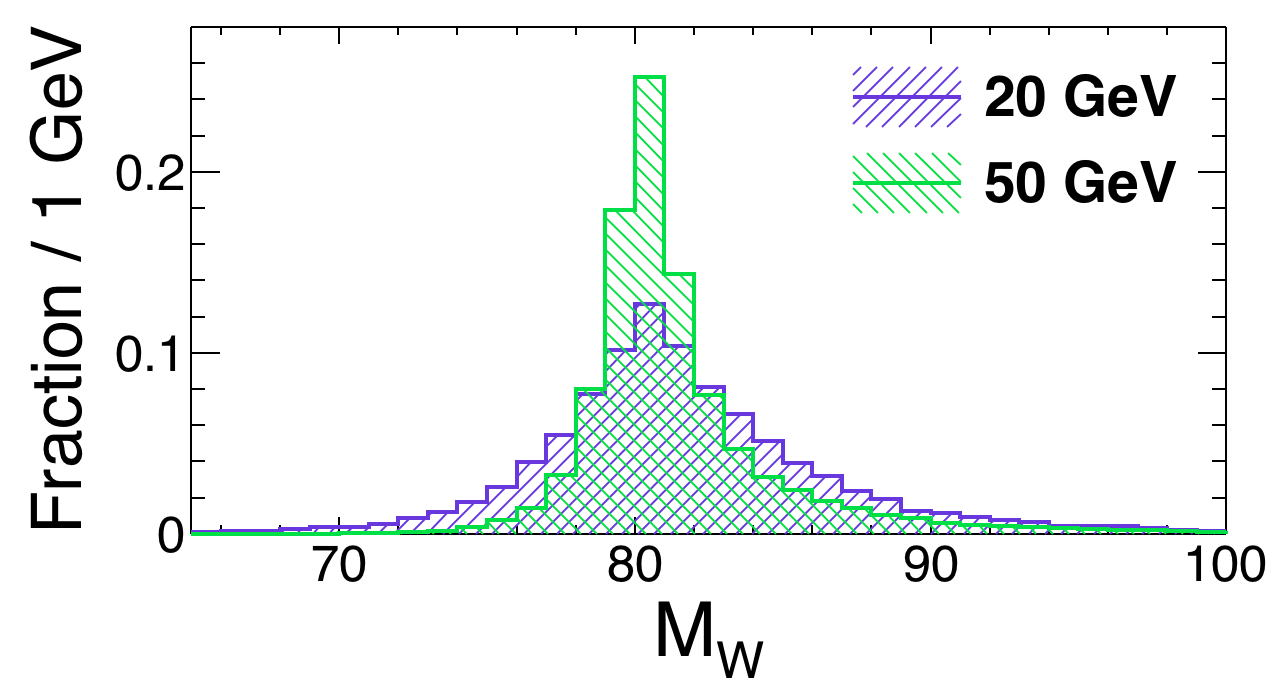}
\caption{Distributions of the reconstructed N mass $M_N$ (left) and W mass $M_W$ (right) for Dirac signals with $m_N$ = 20 and 50 GeV after applying the basic cuts and b-jets veto.}
\label{fig:recmNmW}
\end{figure}

The distributions of the reconstructed $W$ and $N$ masses for Dirac signals using this method, after applying the basic cuts and b-jet vetoes, are shown in Fig.~\ref{fig:recmNmW}.
Since usually the $N$ mass cannot be reconstructed correctly with the incorrect neutrino longitudinal momentum $p_{z,\nu}$ or with the wrong lepton  from the same-sign lepton pair ($l^{\pm} l^{\pm}$), and a good reconstructed $W$ mass also requires a correct $p_{z,\nu}$,
the sharp resonances around our benchmarks $m_N=$ 20 GeV and 50 GeV for the $M_{N}$ distributions, and around $m_W= 80.5$ GeV for the $M_{W}$ distribution, indicate that one can indeed find the correct $p_{z,\nu}$ and identify the lepton from the $N$ decay effectively by minimizing the $\chi^2$ of Eq.~(\ref{eqn:Chi2}). Since the leptons from the $N$ decay for $m_N= 20$ GeV case are softer and affected more by the lepton threshold cuts than for $m_N= 50$ GeV case, the corresponding peaks in the $M_{W}$ distribution for the former case are less sharp compared to the $m_N= 50$ GeV case.

\begin{figure}[h]
\includegraphics[scale=0.09]{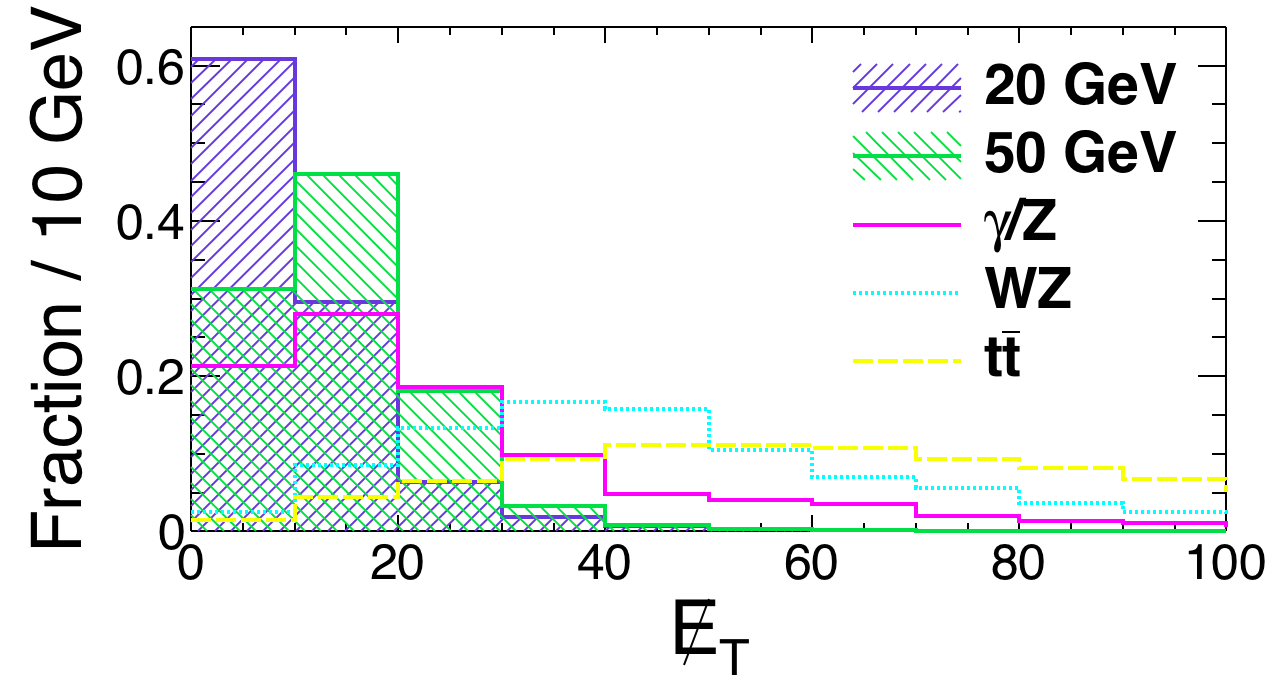}
\includegraphics[scale=0.09]{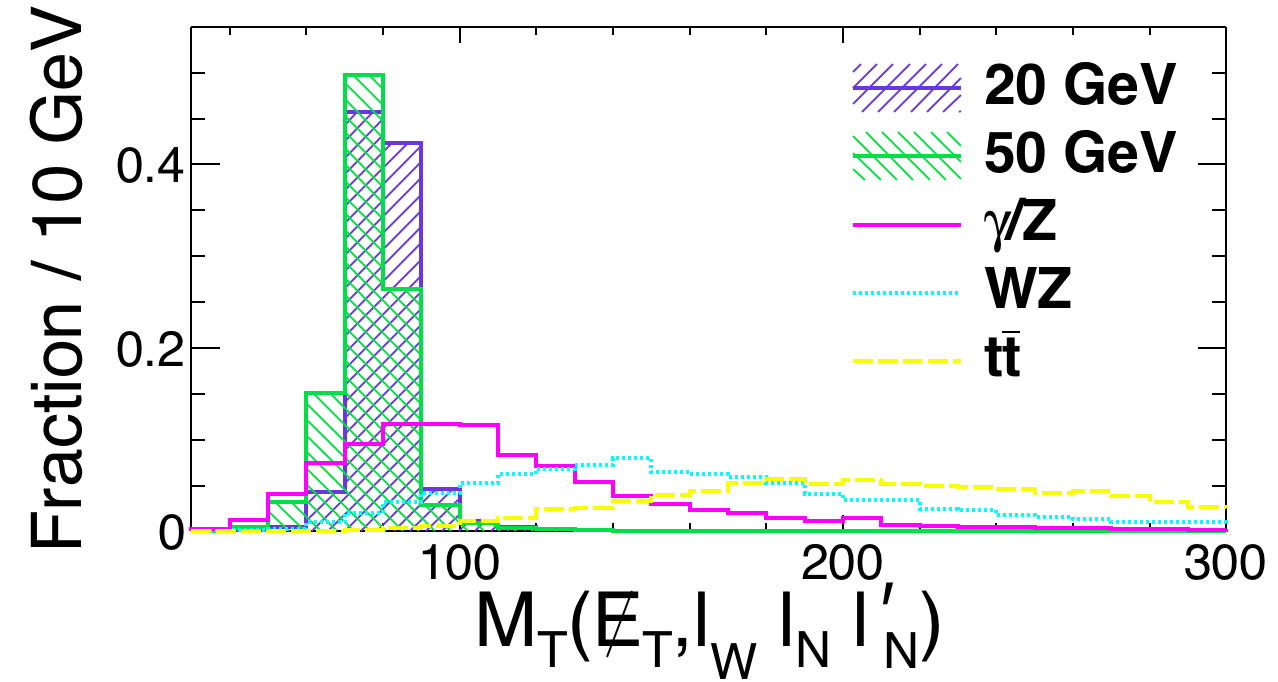}
\includegraphics[scale=0.09]{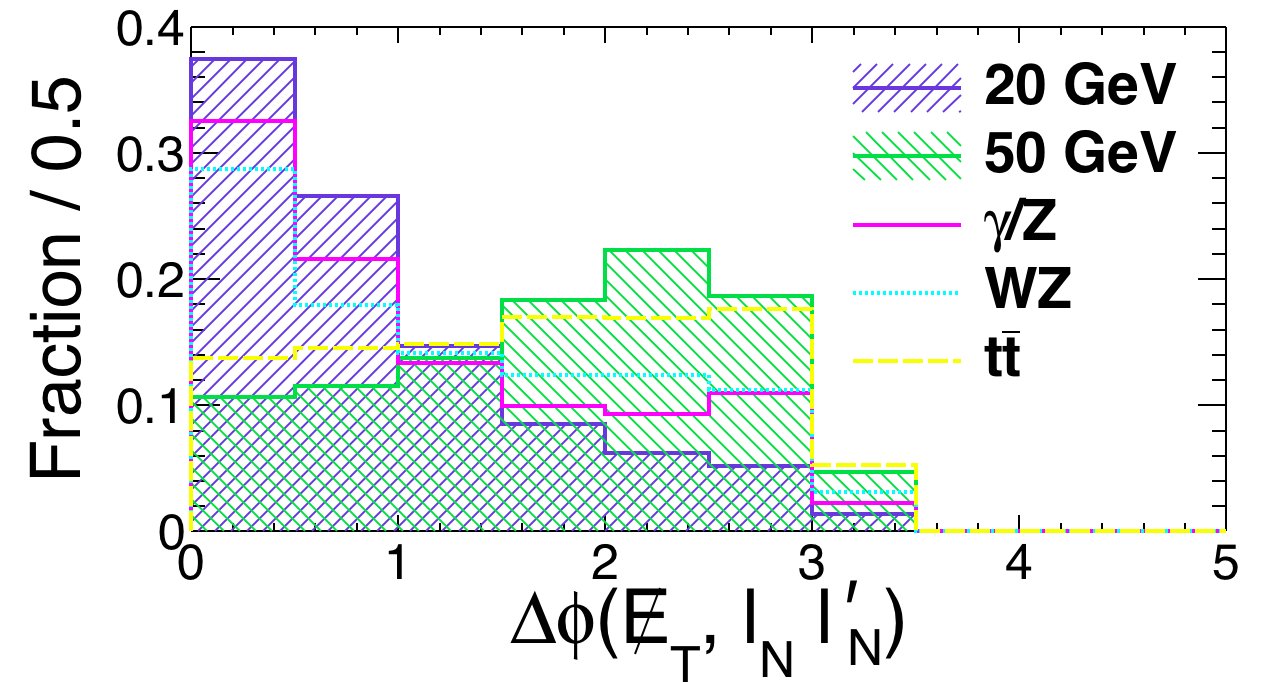}
\includegraphics[scale=0.09]{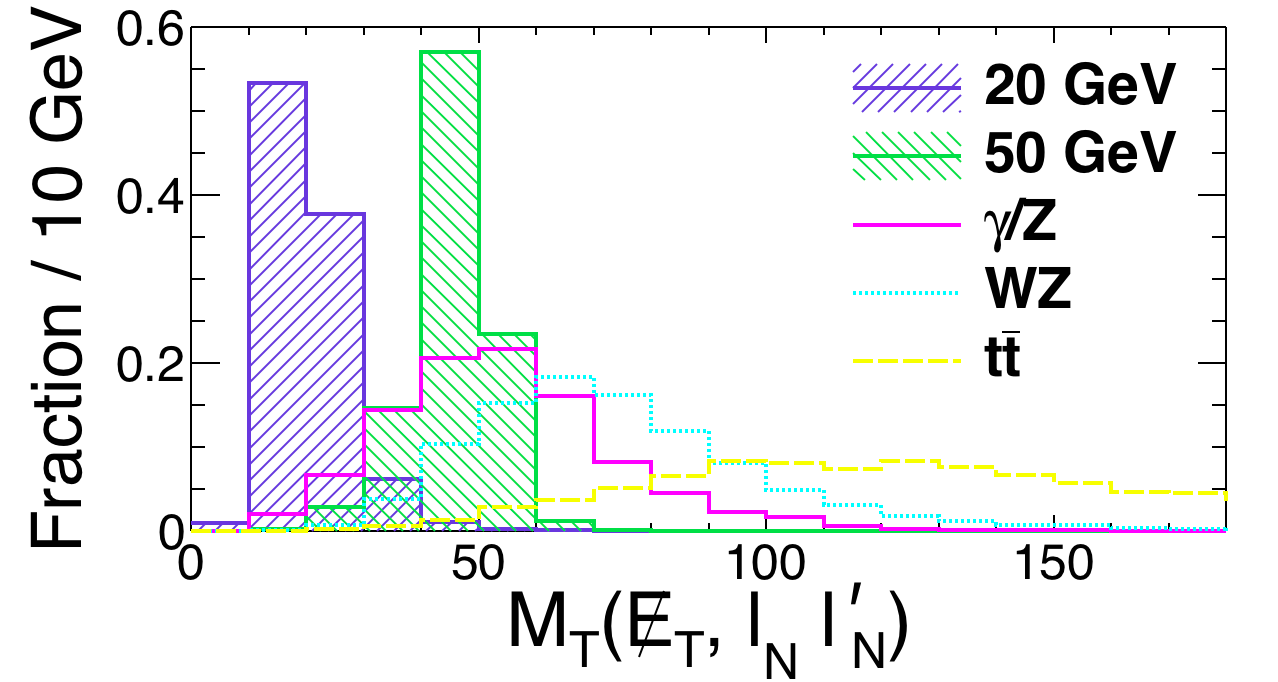}
\includegraphics[scale=0.09]{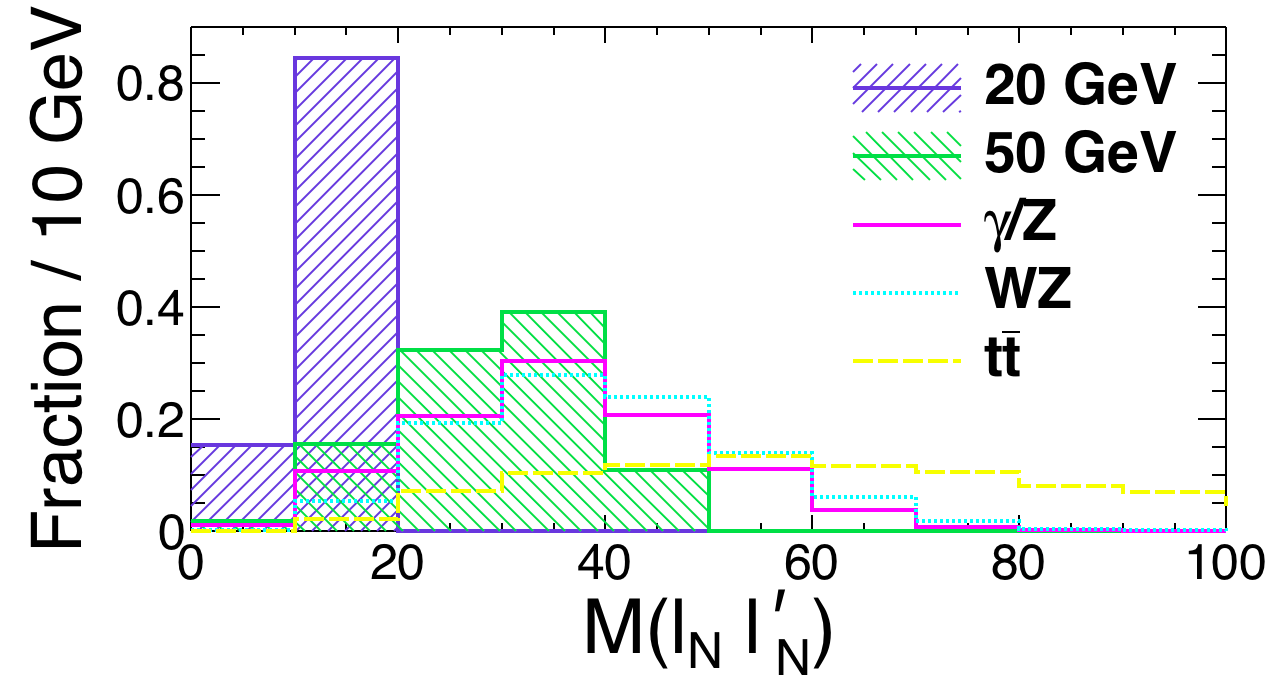}
\includegraphics[scale=0.09]{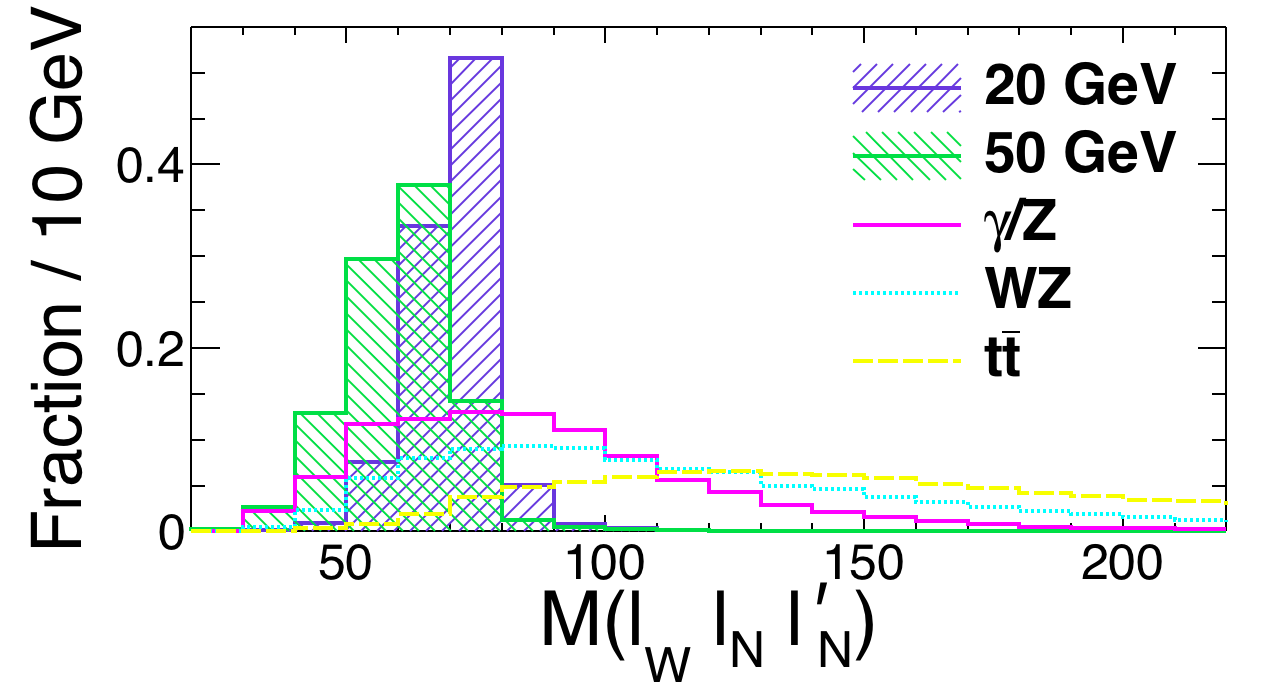}
\caption{Kinematial distributions for Dirac signals with $m_N$ = 20 and 50 GeV, and SM backgrounds of $\gamma^*/Z$+jets, $WZ$+jets and $t\bar{t}$+jets after applying the basic cuts and b-jets veto.}
\label{fig:DistrDrcVSbg}
\end{figure}

Once the correct neutrino longitudinal momentum $p_{z,\nu}$ is found and the right lepton $l^{\pm}$ from the $N$ decay is identified from the lepton pair $l^{\pm} l^{\pm}$ by minimizing the above $\chi^2$ of Eq.~(\ref{eqn:Chi2}), a set of different kinematical observables that are sensitive to the presence of a heavy sterile neutrino $N$, and also sensitive to its Dirac/Majorana character, can be constructed. The full list of these observables will be stated in Section~\ref{sec:NvsBG}.
In Fig.~\ref{fig:DistrDrcVSbg}, we present the distributions of some of them for both Dirac $N$ signals with $m_N$ = 20 and 50 GeV, and for the SM backgrounds of $\gamma^*/Z$, $WZ$ and $t\bar{t}$ after applying the basic cuts and the b-jet vetoes.
One can see that the distributions for signals and backgrounds are quite different, thus these observables can actually be used to reduce the SM backgrounds effectively. Compared with the $m_N$ = 50 GeV case, most signal distributions for 
$m_N= 20$ GeV are more separate from the backgrounds, so these observables can be more efficient to reject the backgrounds and can lead to larger significances for the $m_N= 20$ GeV benchmark point.


\section{Discovering sterile neutrinos with trilepton modes}
\label{sec:NvsBG}

In this section we describe our strategy to search for heavy sterile neutrinos using trileptons at the LHC.
After applying the basic cuts and b-jet vetoes, a Multi-Variate Analysis (MVA) is performed to exploit useful observables and maximally reduce the SM background.
We use the Boosted Decision Trees (BDT) method in the \emph{Toolkit for MultiVariate data Analysis} (TMVA) package~\cite{TMVA2007}, and input 
the following kinematical observables ($i$)-($viii$), which include those presented in Fig.~\ref{fig:DistrDrcVSbg}, for training and test processes:\\
($i$) the missing energy $\met$;\\
($ii$) the scalar sum of $p_T$ of all jets $H_T$; \\
($iii$) the invariant mass of the system of leptons 
$M(l_W l_N {l^\prime}_N)$, 
$M(l_W l_N)$, 
$M(l_W {l^\prime}_N)$, 
$M(l_N {l^\prime}_N)$;\\
($iv$) the azimuthal angle difference $\Delta\phi$ between two leptons 
$\Delta\phi(l_W, {l^\prime}_N)$, 
$\Delta\phi(l_N, {l^\prime}_N)$;\\
($v$) the transverse mass $M_T$ of the system formed by the missing momentum plus lepton(s) 
$M_T(\met, l_W)$,
$M_T(\met, l_N {l^\prime}_N)$, 
$M_T(\met, l_W l_N {l^\prime}_N)$;\\
($vi$) the azimuthal angle difference $\Delta\phi$ between the missing transverse momentum and lepton(s) 
$\Delta\phi(\met, l_N {l^\prime}_N)$, 
$\Delta\phi(\met, l_W)$;\\
($vii$) the transverse mass $M_T$ of the system formed by the missing momentum plus lepton(s) $M_T(\met, l_N)$, $M_T(\met, {l^\prime}_N)$, and $M_T(\met, {l^\prime}_N l_W )$;\\
($viii$) the azimuthal angle difference $\Delta\phi$ between the missing transverse momentum and lepton(s) $\Delta\phi(\met, l_N)$, 
$\Delta\phi(\met, {l^\prime}_N)$, and $\Delta\phi(\met, {l^\prime}_N l_W )$.\\
The observables ($vii$) and ($viii$) in particular are found to differ between the LNC and LNV processes and can be utilized to determine the Majorana/Dirac nature of $N$. That part of the study is presented in Section \ref{sec:drcVSmaj}.

For the Dirac (Majorana) sterile neutrinos, the simulation data of the LNC (LNC plus LNV) processes are inputs as the signal sample, while the total SM background data ($\gamma^*/Z$, WZ, and $t\bar{t}$ inclusively) are inputs as the background sample for the TMVA training and test processes.

\begin{figure}[h]
\includegraphics[scale=0.09]{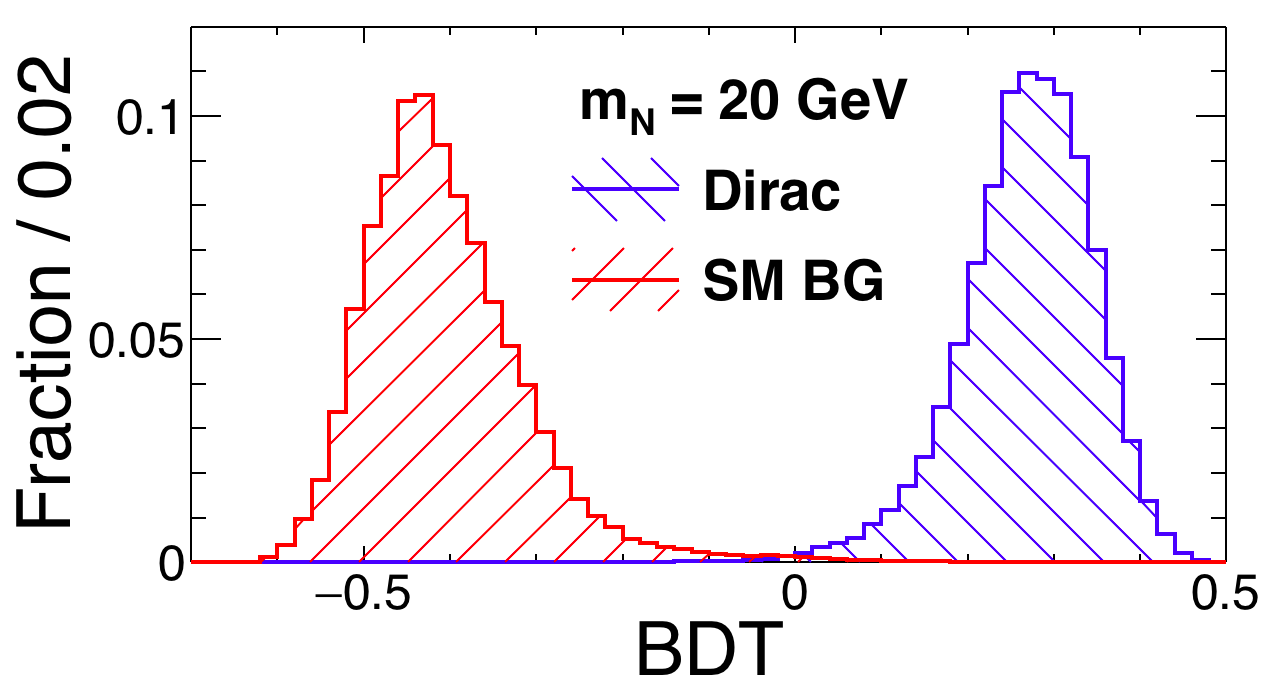}
\includegraphics[scale=0.09]{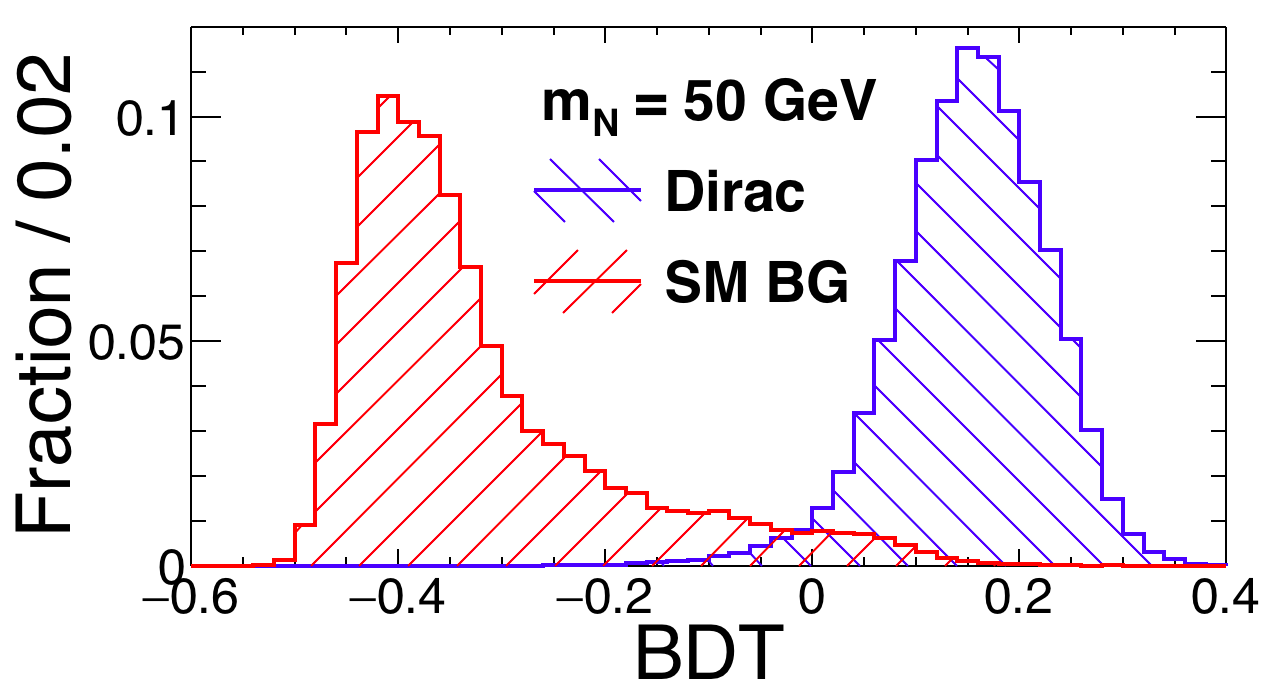}
\caption{ Distributions of BDT response for Dirac signal (blue) with $m_N$ = 20 (left) and 50 (right) GeV, and total SM backgrounds (red) including $\gamma^*/Z$+jets, WZ+jets and $t\bar{t}$+jets.}
\label{fig:BDTdrcVSbg}
\end{figure}

Fig.~\ref{fig:BDTdrcVSbg} shows the distributions of the BDT response
for the Dirac sterile neutrino signals and the total SM background including $\gamma^*/Z$+jets, WZ+jets and $t\bar{t}$+jets, in the two benchmark cases. The kinematical distributions of the signal and of the SM backgrounds differ from each other more for $m_N = $ 20 GeV than for $m_N = $ 50 GeV.

In Table~\ref{tab:sigVSbg_mN20} we show the number of events for both Dirac and Majorana signals and the SM backgrounds at the 14 TeV LHC with integrated luminosity of $3000~\mathrm{fb}^{-1}$, for the case with $m_N = 20$ GeV. The first two rows show the number of events after basic cuts and b-jet vetoes. The number of events using the Cut-and-Count (CC) method used in Ref.~\citep{Dib:2016wge} are shown in the third row. The numbers of events for Dirac (Majorana) sterile neutrinos using the BDT method are shown in the fourth (fifth) row.
For our benchmark point $m_N$ = 20 GeV Dirac (Majorana) sterile neutrino, one can get a statistical significance $\mathit{SS} = N_s/\sqrt{N_s+N_b}$ of about 2.6 (5.8) for the CC method and of about 6.6 (10.7) for the BDT method, where $N_s$ is the number of signal events, while $N_b$ is the corresponding number of total SM background events.
Similarly, for the benchmark point $m_N = 50$ GeV the numbers are shown in Table~\ref{tab:sigVSbg_mN50}, where we find significances of about 2.3 (4.8) for the CC method and of about 5.1 (9.0) for the BDT method.
From Fig.~\ref{fig:BDTdrcVSbg}, one can see that the BDT cut is more efficient to reject the SM backgrounds for the $m_N = 20$ GeV than for 50 GeV, thus higher significances can be expected for the $m_N = 20$ GeV benchmark point. This is indeed what is found by comparing Tables~\ref{tab:sigVSbg_mN20} and \ref{tab:sigVSbg_mN50}. 

\begin{table}[h]
\centering
\begin{tabular}{ccccccc}
\hline
\hline
Cuts            & Dirac & Majorana & $\gamma^*/Z$ & WZ & $t\bar{t}$ & $\mathit{SS}$   \\
Basic cuts         & 54.0 & 133.2 & 4220   & 2658   & 68588   \\
N(b-jets)=0        & 53.1 & 131.1 & 4063.0 & 2497.1 & 31953.5 \\
\hline
CC                 & 44.2 & 110.9 &  209.8 &   25.3 &    16.9 &  2.6 (5.8)  \\
${\rm BDT}>0.183$ & 46.7 &    -  &    1.9 &    1.3 &     0.0 &  6.6 \\
${\rm BDT}>0.171$ &   -  & 120.7 &    5.1 &    1.7 &     0.8 & 10.7 \\
\hline
\hline
\end{tabular}
\caption{Cut flow for signal and background processes with $m_N^{} = 20~\mathrm{GeV}$. Numbers of events correspond to an integrated luminosity of $3000~\mathrm{fb}^{-1}$ at the $14~\mathrm{TeV}$ LHC.}
\label{tab:sigVSbg_mN20}
\end{table}

\begin{table}[h]
\centering
\begin{tabular}{ccccccc}
\hline
\hline
Cuts            & Dirac & Majorana & $\gamma^*/Z$ & WZ & $t\bar{t}$ & $\mathit{SS}$ \\
Basic cuts         & 108.4 & 228.8 & 4220   & 2658   & 68588 &   \\
N(b-jets)=0        & 106.7 & 225.2 & 4063.0 & 2497.1 & 31953.5 & \\
\hline
CC                 &  91.9 & 193.9 & 1283.1 &  120.7 &    48.9 & 2.3 (4.8) \\
${\rm BDT}>0.138$ &  64.4 &    -  &   25.7 &   47.5 &    21.1 & 5.1 \\
${\rm BDT}>0.138$ &    -  & 143.2 &   31.0 &   52.8 &    27.0 & 9.0 \\
\hline
\hline
\end{tabular}
\caption{Cut flow for signal and background processes with $m_N^{} = 50~\mathrm{GeV}$. Numbers of events correspond to an integrated luminosity of $3000~\mathrm{fb}^{-1}$ at the $14~\mathrm{TeV}$ LHC.}
\label{tab:sigVSbg_mN50}
\end{table}

\begin{figure}[h]
\includegraphics[scale=0.09]{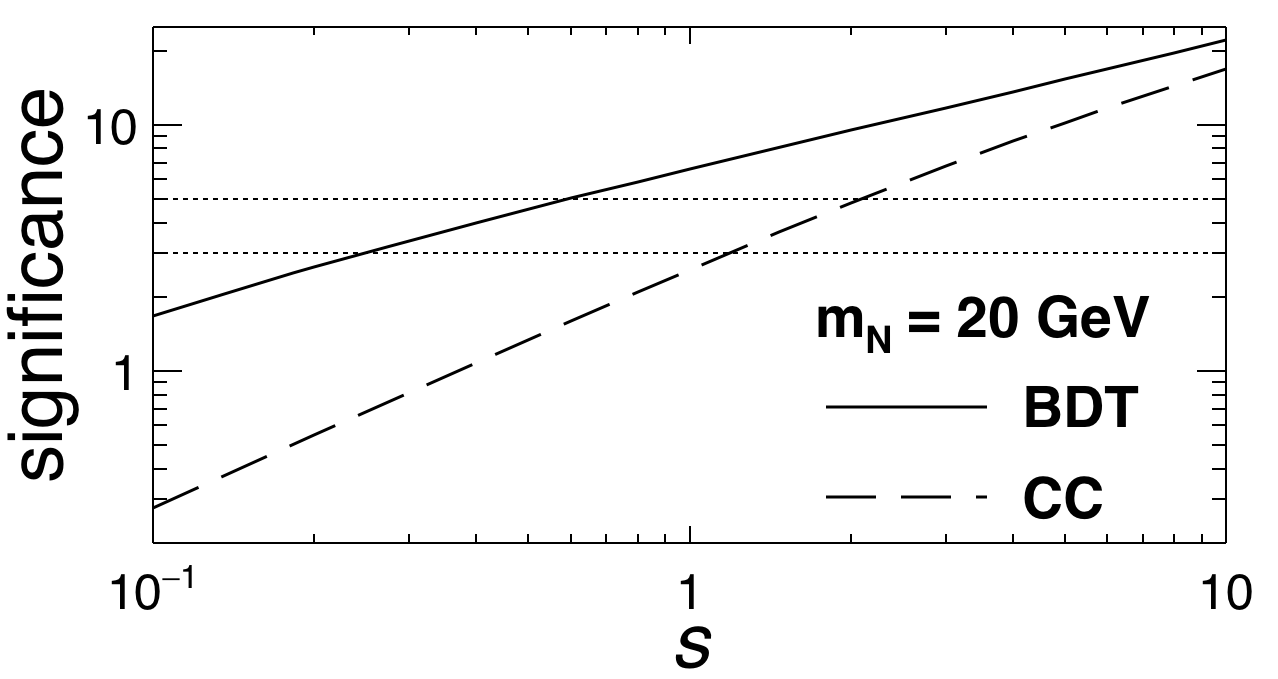}
\includegraphics[scale=0.09]{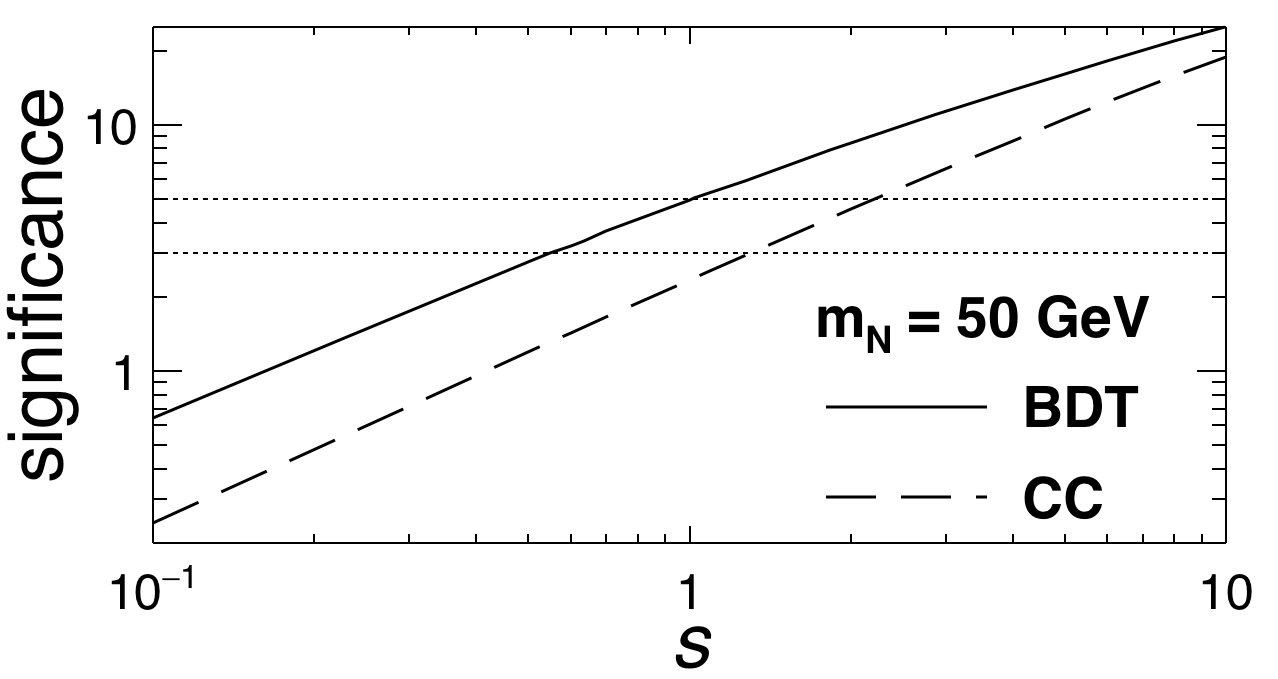}
\caption{Discovery and exclusion limits for Dirac sterile neutrinos with $m_N$ = 20 (left) and 50 (right) GeV. }
\label{fig:sgfDrc}
\end{figure}

Fig.~\ref{fig:sgfDrc} 
shows the discovery and exclusion curves for Dirac sterile neutrinos, for both the BDT and CC methods. According to Table~\ref{tab:Nsig}, for a given $m_N$ the production rates for Dirac sterile neutrinos (LNC processes) depend on the factor $s$ only, so the observables at the LHC in the Dirac scenario can just constrain the parameter $s$. By exploiting more useful kinematical observables and better optimization compared with the CC method, the BDT method can greatly enhance the discovery and exclusion limits. Nevertheless,  the performance of the BDT method becomes close to that of the CC method for small $s$ values, due to the small number of signal events.
Using the BDT method, for the benchmark $m_N = 20$~GeV one can get significances above $5.0\,\sigma\, (3.0\,\sigma)$ when $s \geq 0.55\, (0.25) $, and for the benchmark $m_N = 50$~GeV similar significances are reached when $s \geq 1.02\, (0.55)$.

\begin{figure}[h]
\includegraphics[scale=0.09]{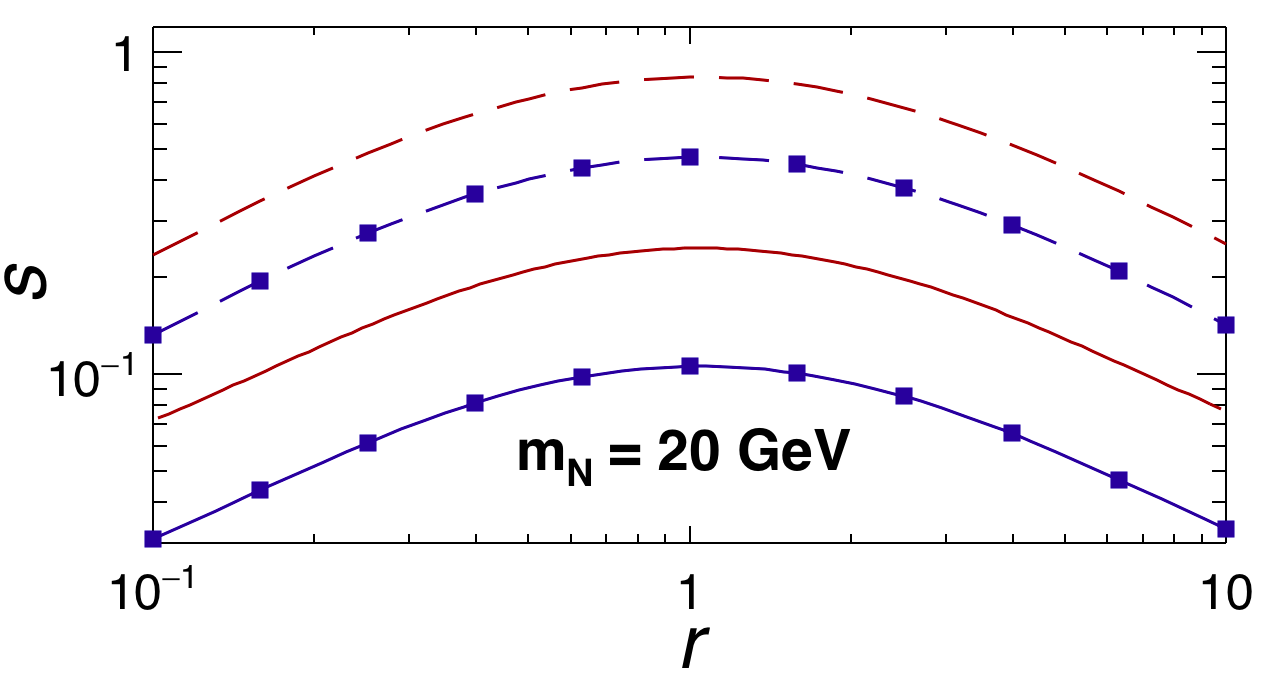}
\includegraphics[scale=0.09]{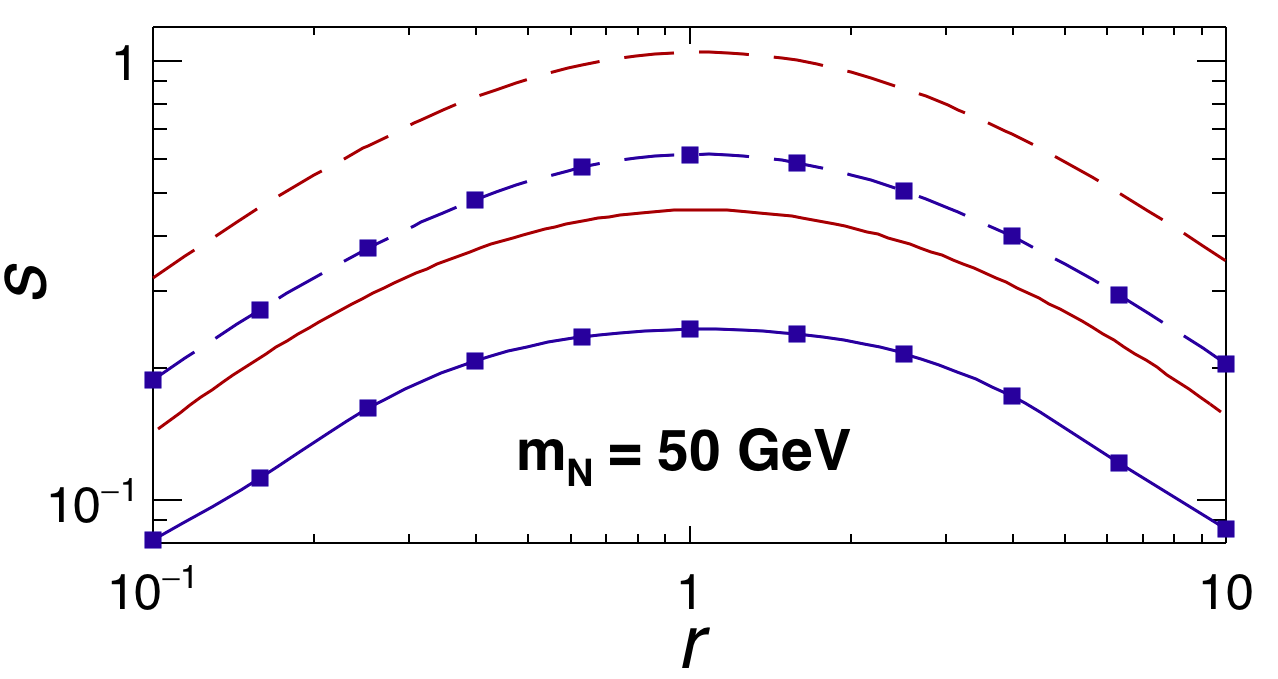}
\caption{Discovery and exclusion limits for Majorana sterile neutrinos with $m_N$ = 20 (left) and 50 (right) GeV. The blue curves marked with squares correspond to 3-$\sigma$ limit, while the red curves correspond to 5-$\sigma$ limit; solid lines for BDT method and dashed lines for CC method.}
\label{fig:sgfMaj}
\end{figure}

Fig.~\ref{fig:sgfMaj} 
shows the discovery and exclusion contour curves for Majorana sterile neutrinos, for both the BDT and CC methods. Since the production rate for Majorana $N$ involves both LNC and LNV processes, it depends on both the normalization $s$ and the ratio $r$ (see Table~\ref{tab:Nsig}). Thus the observables at the LHC in the Majorana scenario can be used to constrain both $s$ and $r$.  Using the BDT method, when $r = 1$ one can get  significances above $ 5.0\,\sigma\,  (3.0\,\sigma)$ with $s \geq 0.24\, (0.11) $ in the case 
$m_N = $ 20 GeV, and with $s \geq 0.46\, (0.25) $ in the case $m_N = $ 50 GeV. For a given value of  $s$,  the significance becomes larger when either $r \gg 1$ or  $r \ll 1$, due to the larger number of signal events. For example, when $r \approx$ 10, one can get significances above $ 5.0\,\sigma\,  (3.0\,\sigma)$  with $s \geq 0.08\, (0.03) $ in the case $m_N = $ 20 GeV, and with $s \geq 0.16\, (0.09) $ in the case $m_N = $ 50 GeV.

\section{Distinguishing between Dirac and Majorana}
\label{sec:drcVSmaj}

\begin{table*}
\begin{ruledtabular}
\begin{tabular}{ccccccccc cccc}
 & \multicolumn{2}{c}{$e^+ e^+ \mu^-$} & \multicolumn{2}{c}{$\mu^+ \mu^+ e^-$} &  \multicolumn{2}{c}{$e^- e^- \mu^+$} & \multicolumn{2}{c}{$\mu^- \mu^- e^+$}
& $l^\pm l^\pm {l^\prime}^\mp$ & $l^+l^+{l^\prime}^-$ & $l^-l^-{l^\prime}^+$ & $l^\pm l^\pm {l^\prime}^\mp$  \\
Cuts & LNC & LNV & LNC & LNV & LNC & LNV & LNC & LNV
& $\gamma^*/Z$ & $W^+ Z$ & $W^- Z$ & $t\bar{t}$ \\
\cline{2-3} \cline{4-5} \cline{6-7} \cline{8-9} \cline{10-10} \cline{11-12} \cline{13-13}
Basic cuts            & 13.6 & 19.5 & 15.0 & 22.0 & 12.1 & 18.2 & 13.3 & 19.5 & 1055.0 & 779.0 & 550.0 & 17147.0 \\
N(b-jets)=0           & 13.4 & 19.2 & 14.7 & 21.7 & 11.9 & 17.9 & 13.1 & 19.2 & 1015.8 & 731.8 & 516.7 &  7988.4 \\
${\rm BDT1} > 0.171$ & 12.2 & 17.7 & 13.5 & 20.0 &  10.9 & 16.5 & 12.0 & 17.7 &   1.2 &  0.5 &    0.4 &     0.2 \\
\end{tabular}
\end{ruledtabular}
\caption{Cut flow for benchmark point with $m_N^{} = 20~\mathrm{GeV}$ and SM backgrounds. From SM backgrounds, $l$ denotes either $e$ or $\mu$. Numbers of events correspond to an integrated luminosity of $3000~\mathrm{fb}^{-1}$ at the $14~\mathrm{TeV}$ LHC. }
\label{tab:1stBDT}
\end{table*}

\begin{figure}[h]
\includegraphics[scale=0.20]{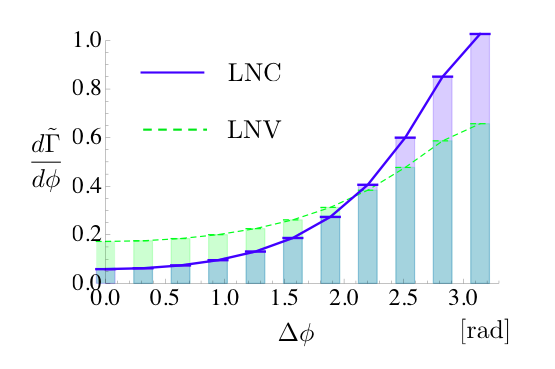}
\includegraphics[scale=0.10]{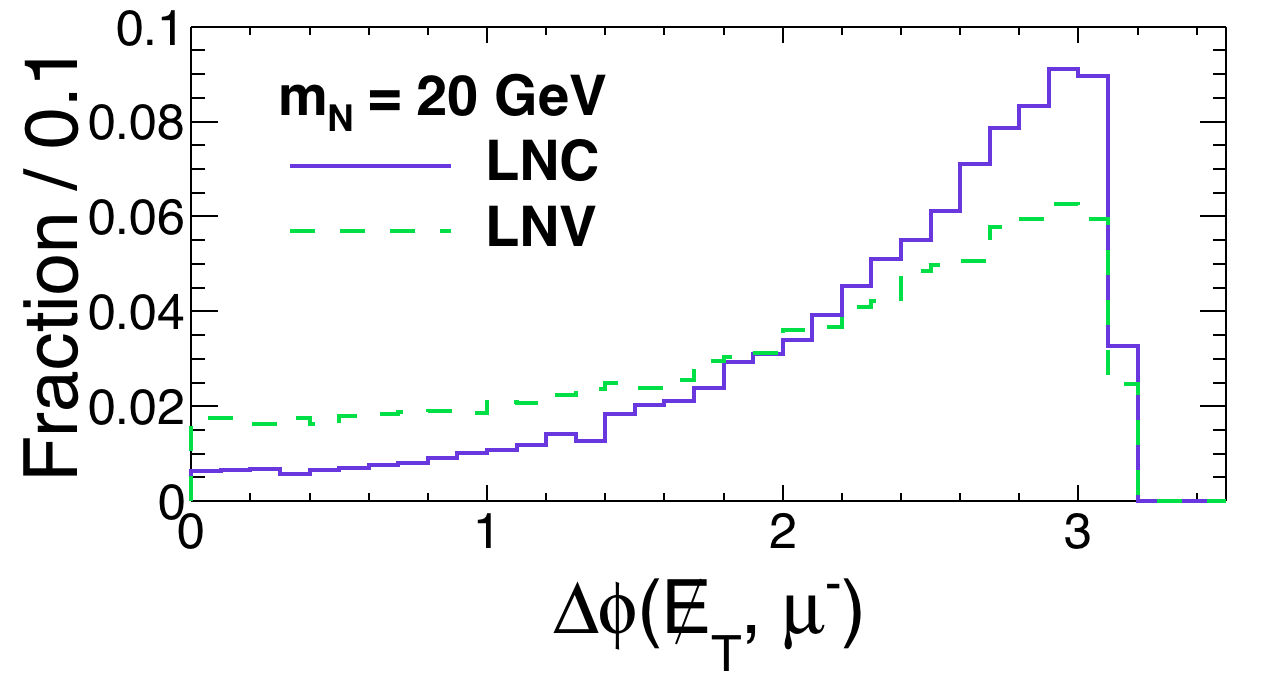}
\caption{
The $\Delta\phi(\met,\mu^-)$ distributions in the $N$ rest frame from the $W^+ \to e^+ e^+ \mu^- \nu$ process for the benchmark point $m_N$ = 20 GeV by theoretical calculation (left) and data simulation (right). Solid blue and dashed green lines correspond to LNC and LNV processes, respectively.
}
\label{fig:obsLNCvsLNVtheory}
\end{figure}

In this section, we show how to distinguish between Dirac and Majorana sterile neutrinos using the trilepton events. 
Recalling that a Majorana $N$ induces both LNV and LNC processes, while a Dirac $N$ induces LNC processes only, a discrimination between Dirac vs. Majorana $N$ can be achieved based on the differences between the LNC and LNV processes.
As mentioned in Section \ref{sec:NvsBG}, among all input observables, the distributions of the observables ($vii$) and ($viii$) are found to differ between the LNC and LNV processes. The corresponding theoretical expressions can be deduced from Section \ref{sec:theoCal}. 
In Fig.~\ref{fig:obsLNCvsLNVtheory} we show, as an example for $m_N = 20$ GeV,
the distribution of $\Delta\phi(\met,l_N^\prime)$  in the $N$ rest frame, where we check that the simulations (right plot) indeed reproduce the distinguishing  features of the theoretical behavior  (left plot), and are not purely an effect of statistical fluctuations.
We use these observables to try to distinguish the Majorana nature of $N$ from a Dirac scenario, the latter taken as the null hypothesis.

In order to exploit the differences between the Dirac and Majorana processes, we must first reduce as much SM background as possible, otherwise the distributions will be dominated by the SM backgrounds and the differences will become imperceptible. 

Therefore, as a first step, after applying the basic cuts and the b-jet vetoes, we perform the first BDT analysis and 
input the kinematical observables ($i$)-($vi$) listed in the first paragraph of Section \ref{sec:NvsBG} to suppress the SM backgrounds.
Simulated Majorana data are input as the signal sample, while the total SM background data are input as the background sample for the TMVA training and testing processes.

Table~\ref{tab:1stBDT} shows the number of events after these cuts for the benchmark case $m_N = 20$ GeV. After the first BDT cut, the total number of events including 
all four different final states ($e^\pm e^\pm\mu^\mp$ and $\mu^\pm\mu^\pm e^\mp$)
for the Dirac signals (the LNC rate only), for the Majorana signals (LNC plus LNV rates) and for the SM backgrounds ($\gamma^*/Z$, $W^{\pm} Z$, and $t\bar{t}$ inclusively), are 48.5, 120.4 and 7.3, respectively. 
The SM backgrounds are reduced to a negligible level, so the sample will be dominated by the $N$-induced signal events.

Since the parameter $s$ is an unknown global scale, the Dirac and Majorana cases cannot be experimentally discriminated purely by the number of events. Therefore, as a second step we adjust $s$ for the Dirac hypothesis 
to a value $s_D$ that matches the number of events for the 
Majorana scenario, 
so that our simulation does not distinguish the two scenarios simply by the rates.

Just as in Ref.~\cite{Dib:2016wge}, the best matched value of $s_{\rm D}$ is found by minimizing the chi-square function
\begin{eqnarray}
\chi^2_{H} = -2\, \underset{s}{\text{min}} \left \{ \text{ln} \left( \prod_i ~ \text{Poiss} \left [ N_i^{\text{expc}}, N_i^{\text{obs}}(s) \right ] \right) \right \},\ \ 
\end{eqnarray}
where $i$ indicates a particular trilepton final state,  Poiss($N^{\rm expc}$, $N^{\rm obs}$) denotes the probability of observing $N^{\rm obs}$ events in Poisson statistics when the number of expected events is $N^{\rm expc}$. 
Here $N^{\rm expc}$ is the expected number of events for 
the Majorana scenario
(LNC + LNV + SM background), while $N^{\rm obs}$ is the observed number of events for the Dirac hypothesis (LNC + SM background).
The best matched parameter $s_{\rm D}$ found in this way gives for the Dirac hypothesis a number of events closest to those of the Majorana case for 
all four different final states ($e^\pm e^\pm\mu^\mp$ and $\mu^\pm\mu^\pm e^\mp$). 
For our $m_N$ = 20 GeV benchmark, the matched parameter $s_{\rm D}$ is found to be around 2.44. After matching, the Dirac hypothesis will have 125.6 events, which is close to the 127.6 events of the Majorana scenario. 

\begin{figure}[h]
\includegraphics[scale=0.09]{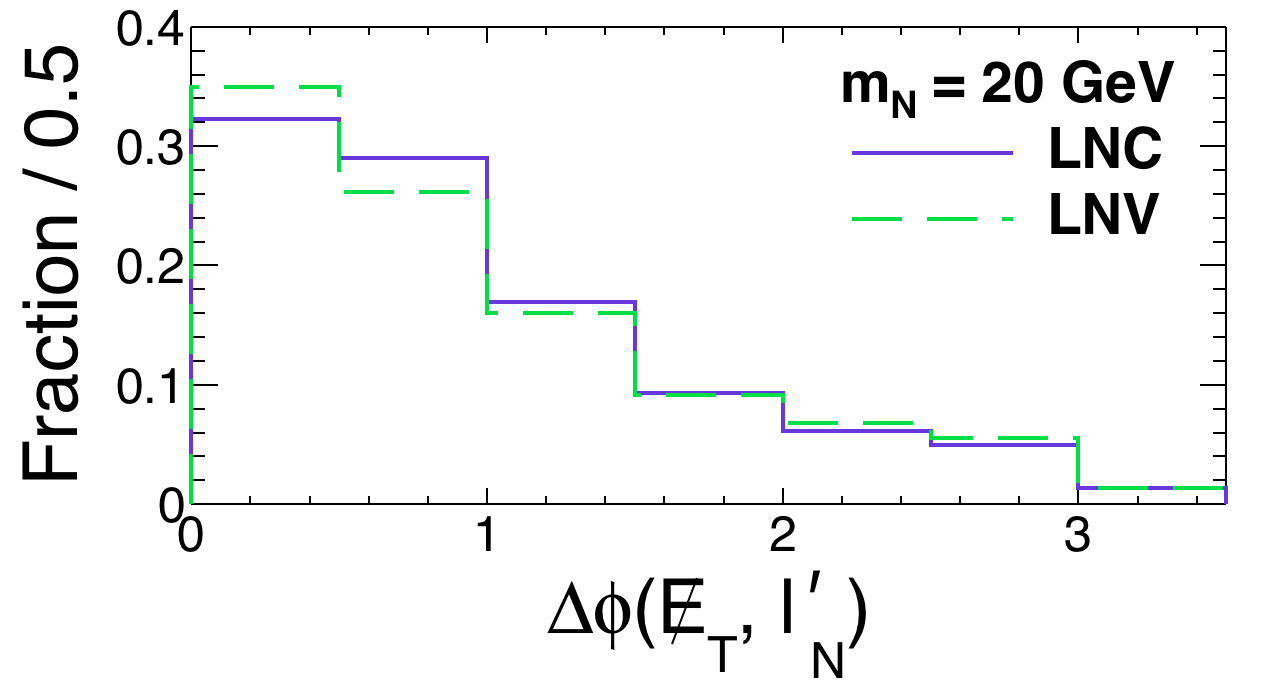}
\includegraphics[scale=0.09]{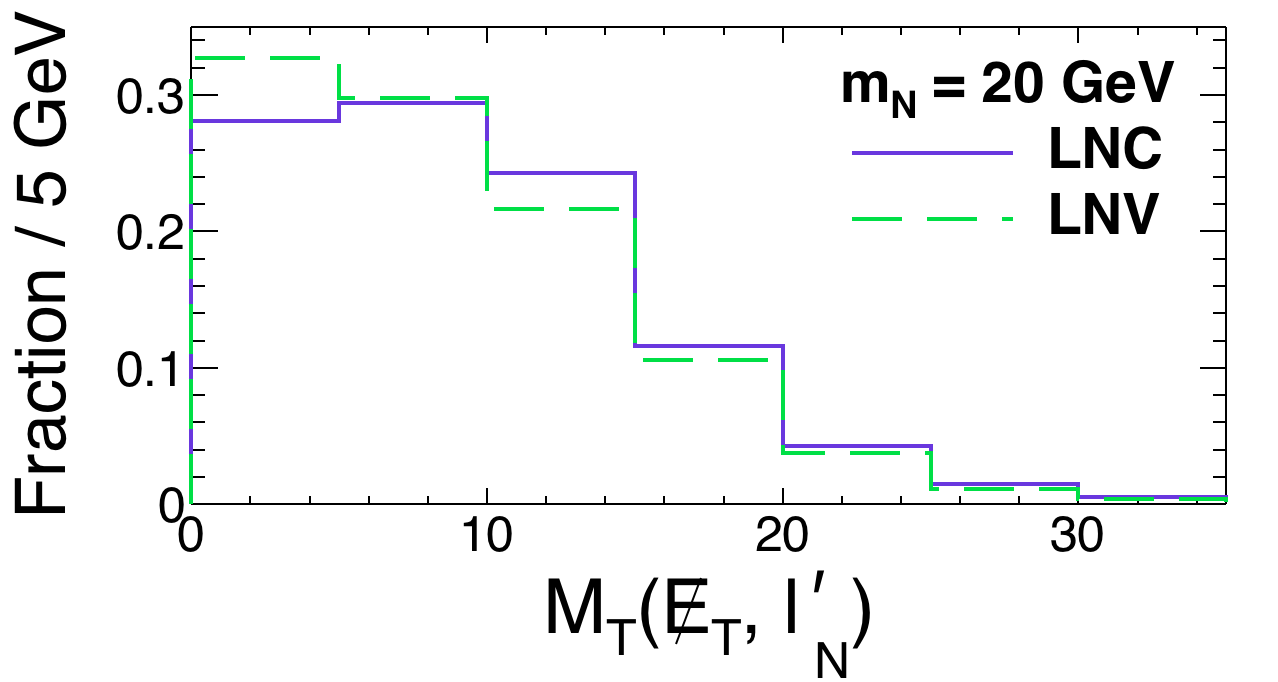}
\includegraphics[scale=0.09]{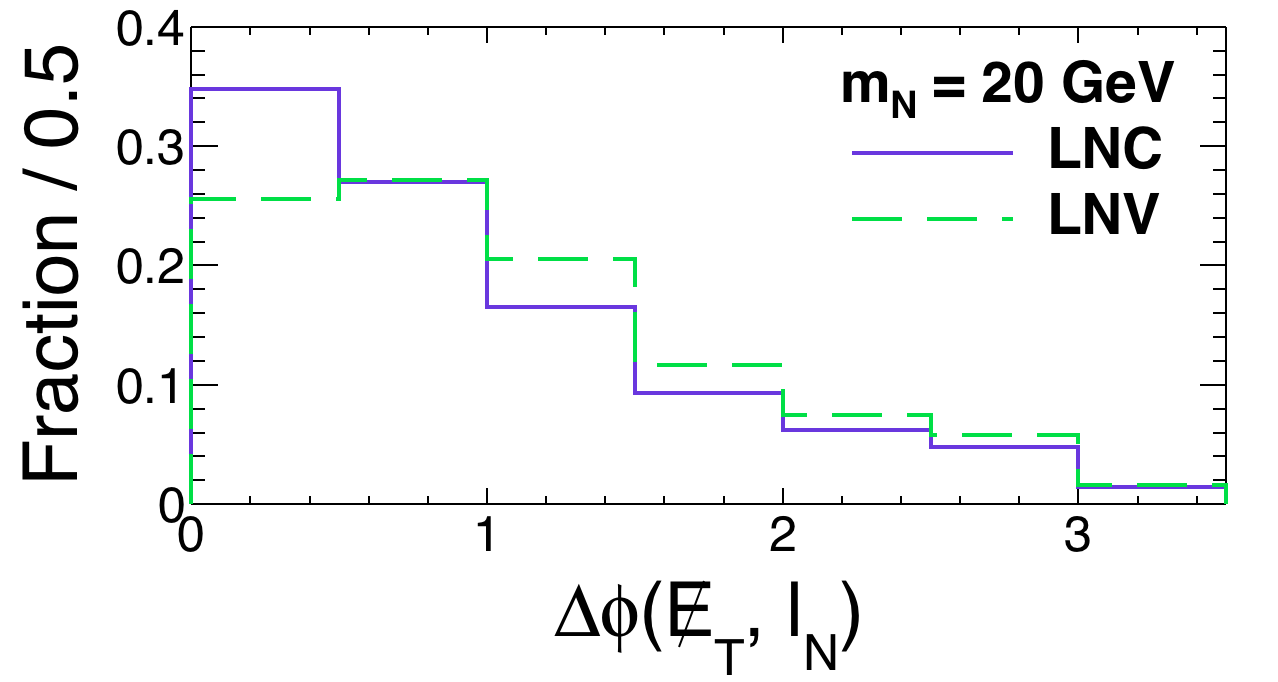}
\includegraphics[scale=0.09]{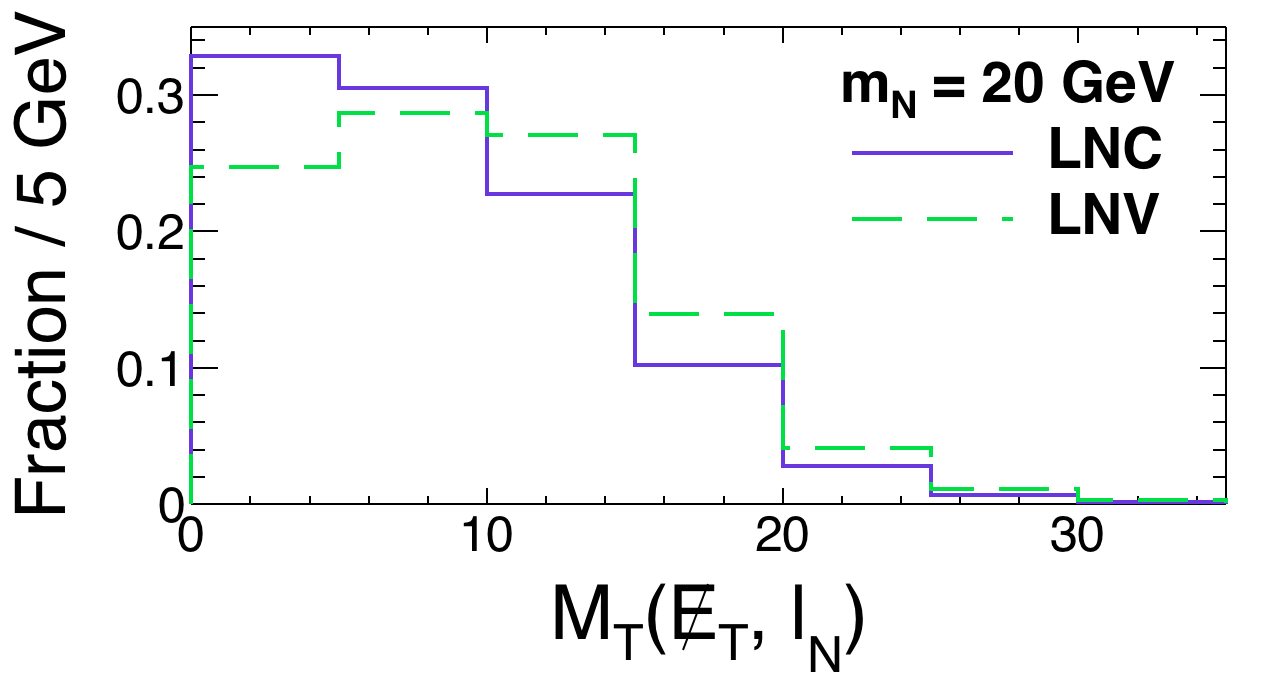}
\includegraphics[scale=0.09]{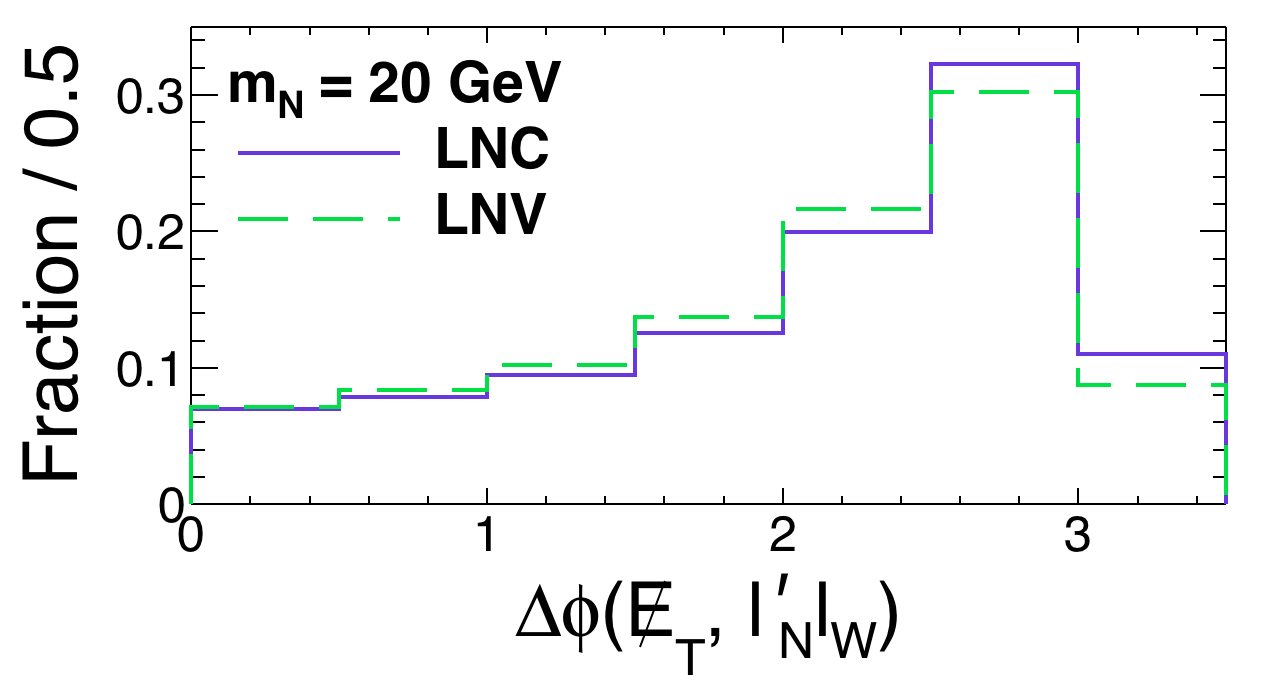}
\includegraphics[scale=0.09]{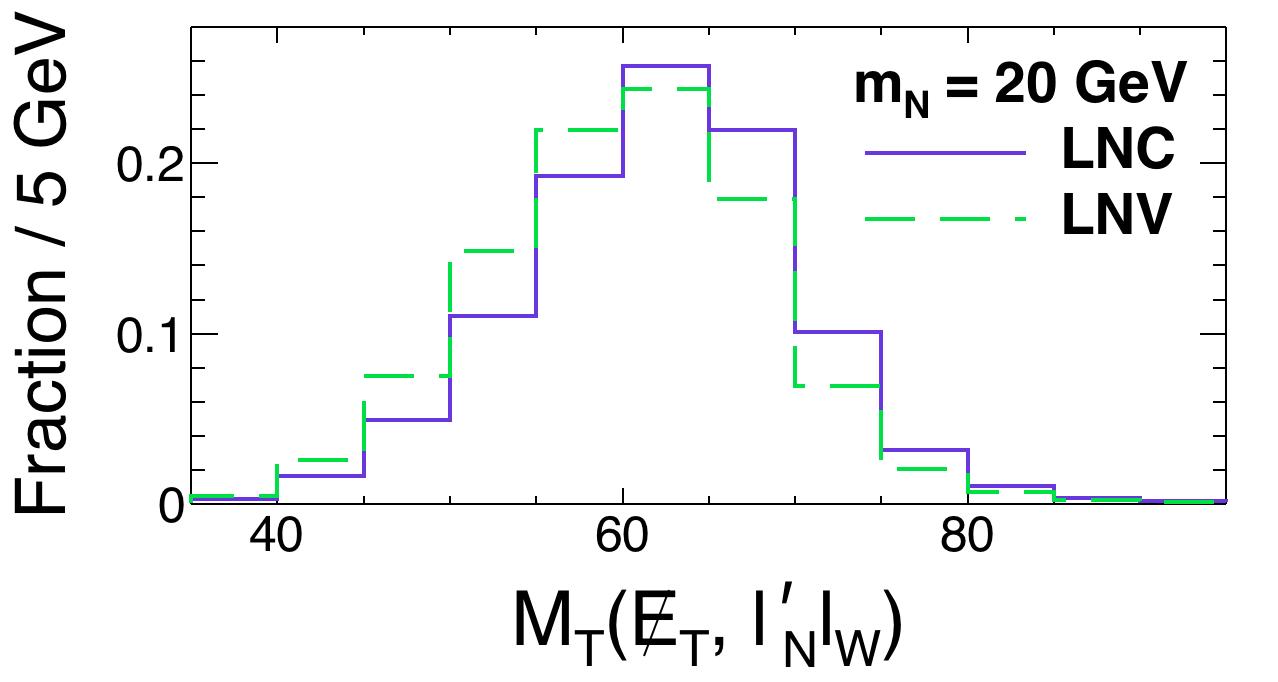}
\caption{Distributions for the benchmark point $m_N$ = 20 GeV after applying the basic cuts, b-jet vetoes and the first BDT cut.}
\label{fig:obsLNCvsLNVmN20}
\end{figure}

\begin{figure}[h]
\includegraphics[scale=0.09]{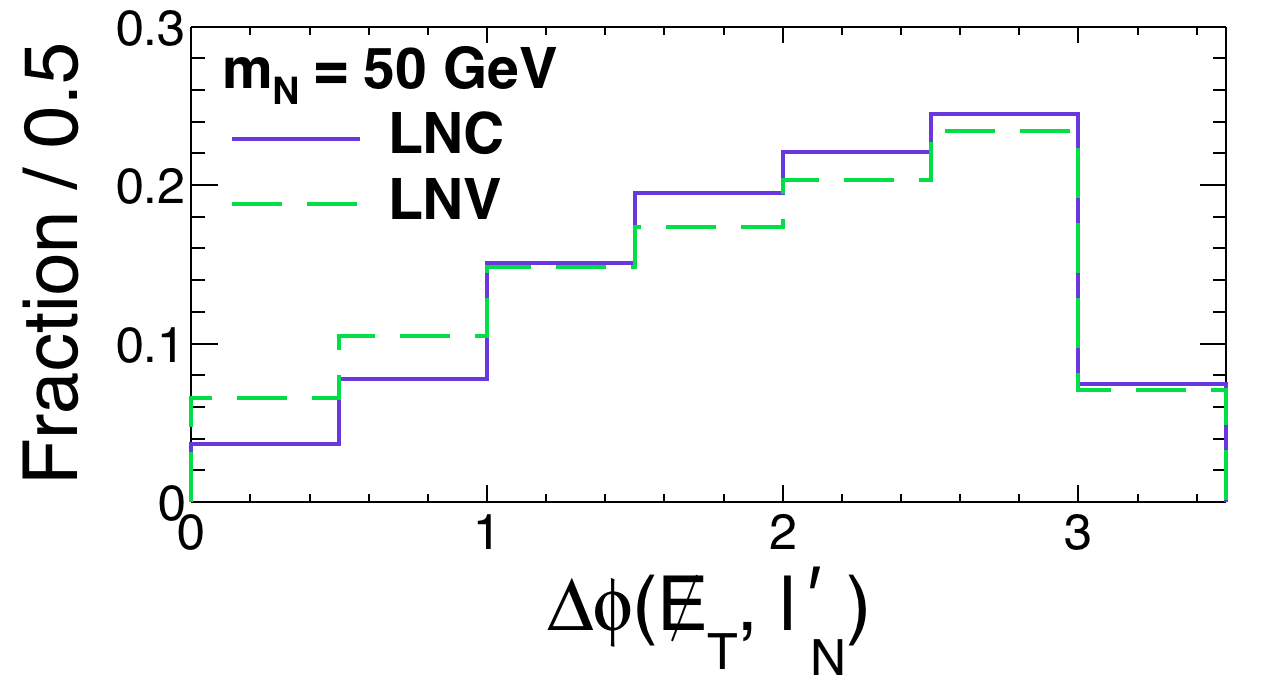}
\includegraphics[scale=0.09]{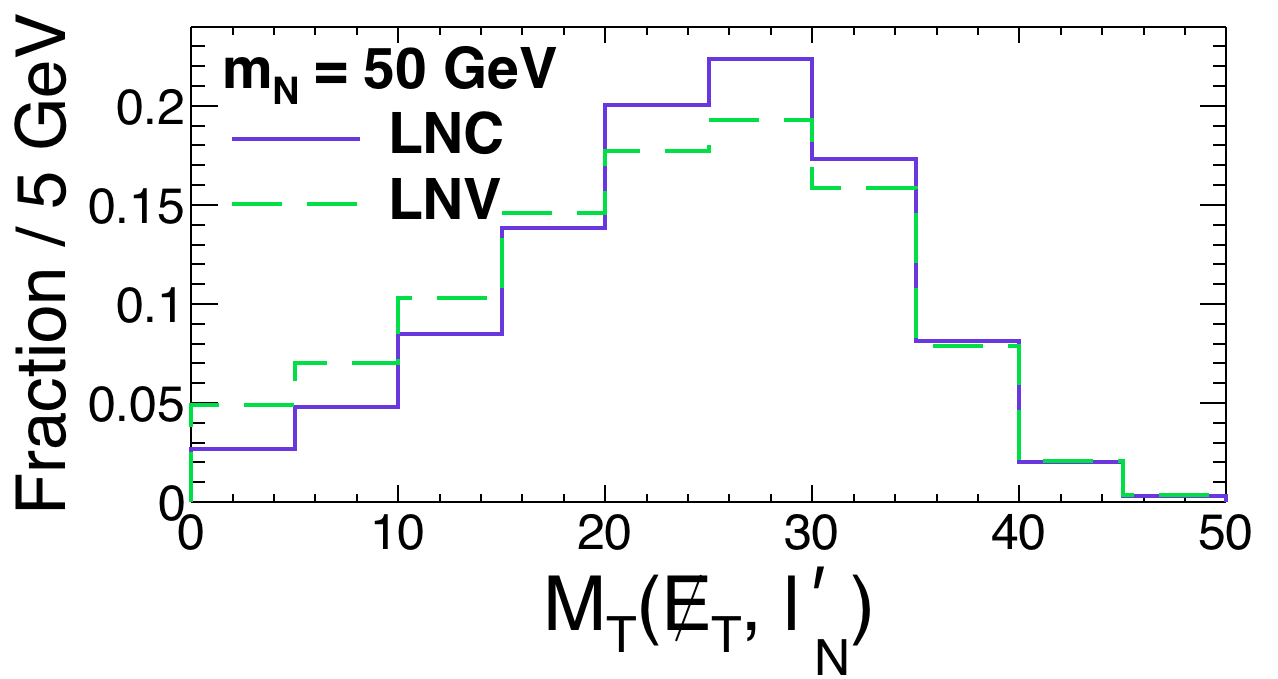}
\includegraphics[scale=0.09]{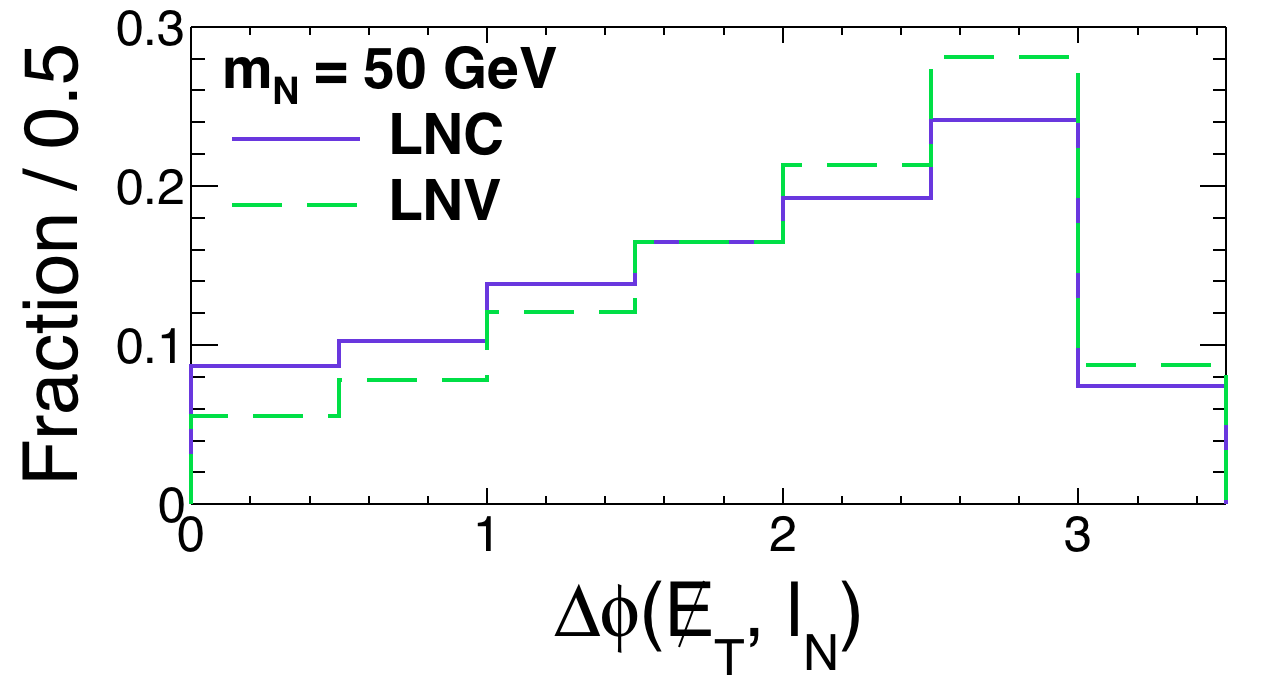}
\includegraphics[scale=0.09]{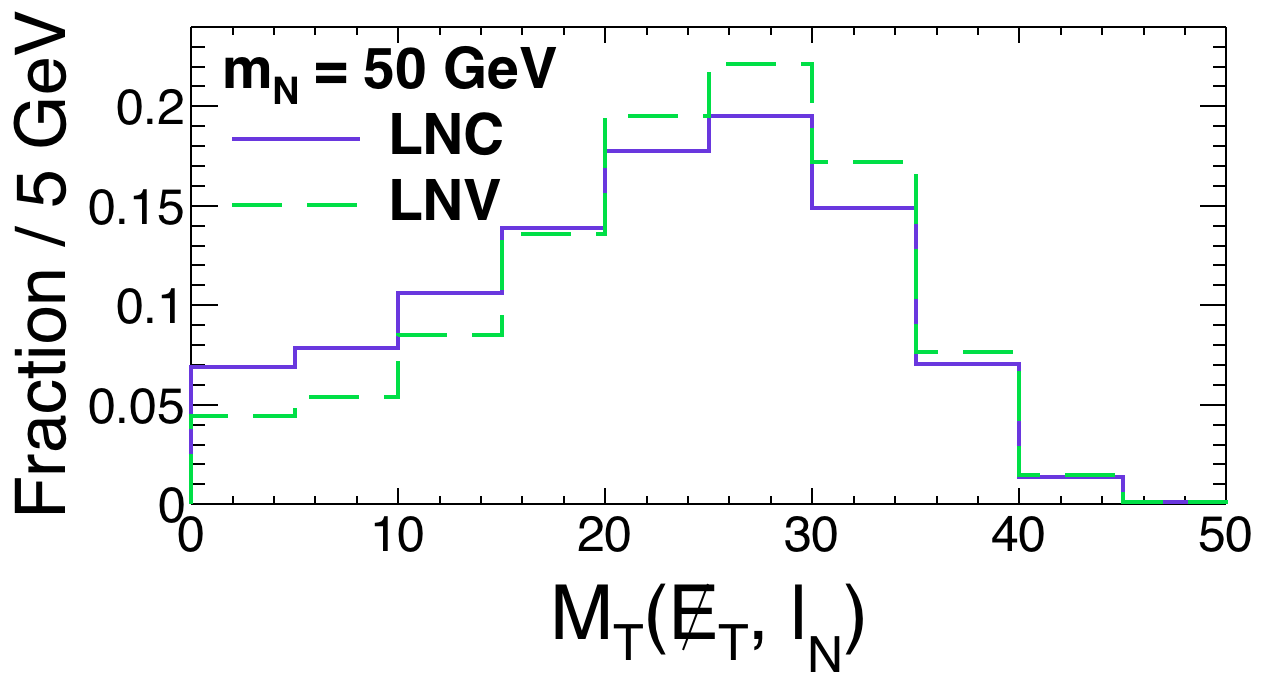}
\includegraphics[scale=0.09]{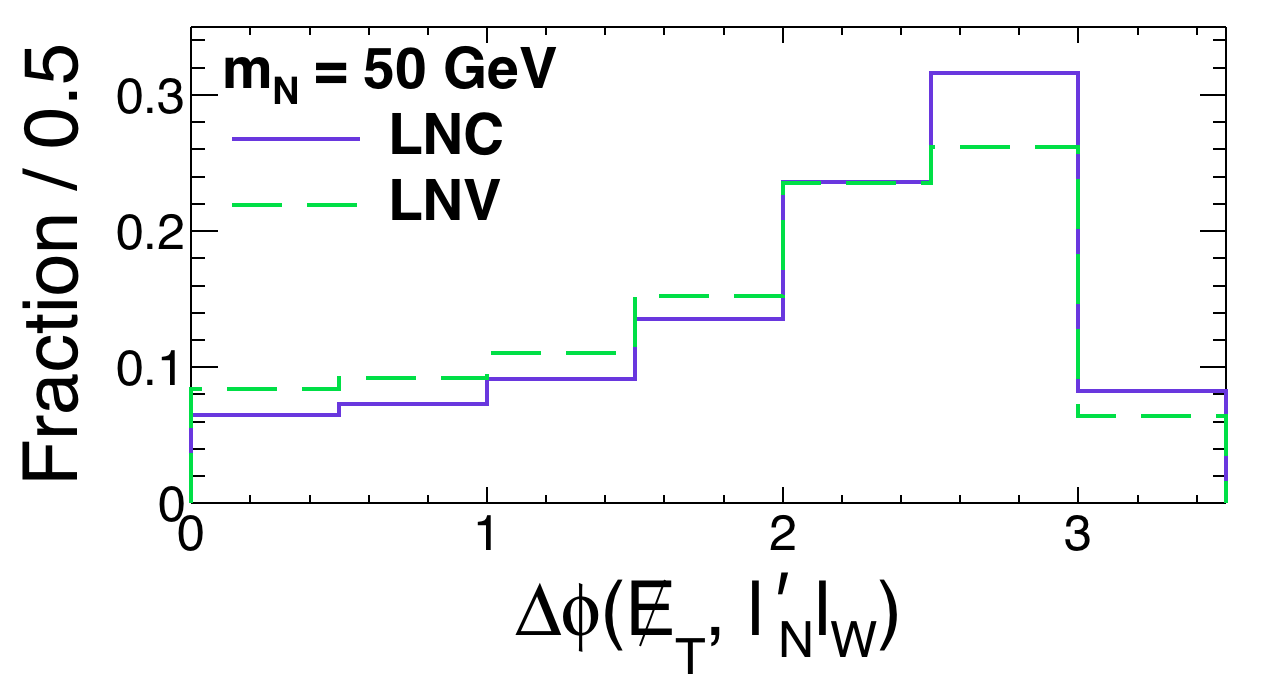}
\includegraphics[scale=0.09]{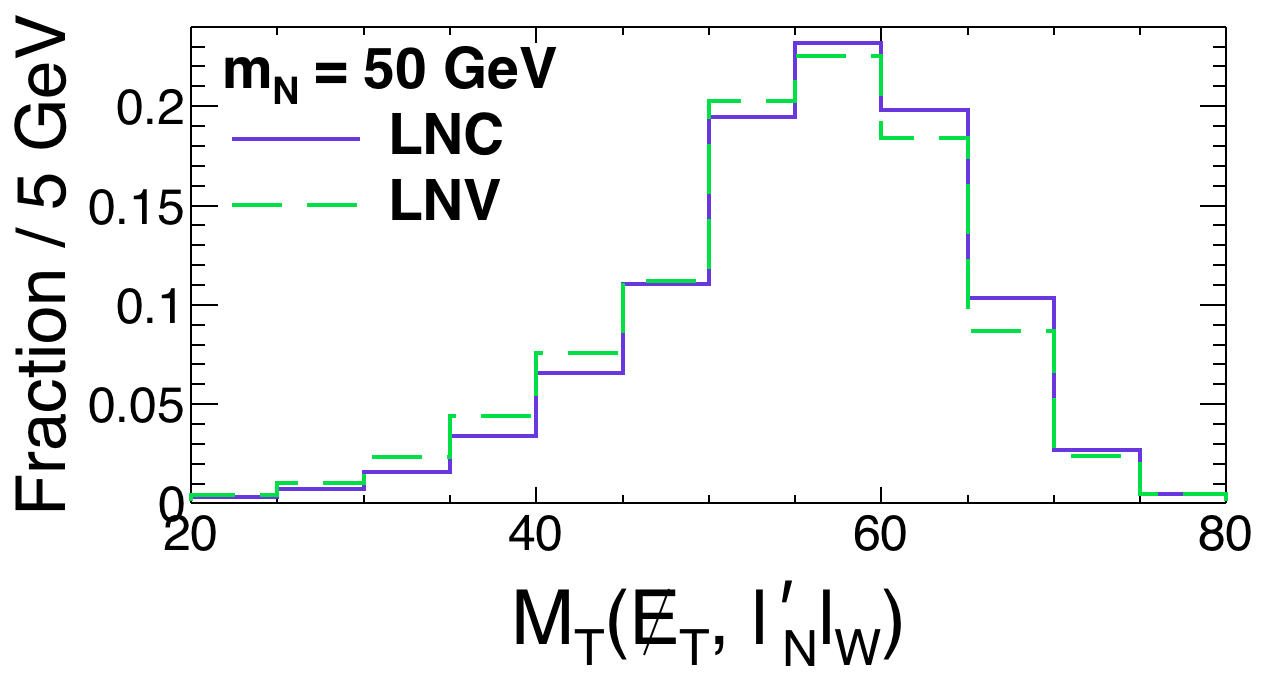}
\caption{Distributions for the benchmark point $m_N$ = 50 GeV after applying the basic cuts, b-jet vetoes and the first BDT cut.}
\label{fig:obsLNCvsLNVmN50}
\end{figure}

As a third step, we perform a second BDT analysis to distinguish a Majorana scenario from the Dirac hypothesis, by exploiting the differences in the kinematical distributions between the LNC and LNV processes.
The input observables are those 
($vii$) and ($viii$) listed in the first paragraph of Section \ref{sec:NvsBG}.
The distributions of these observables for the LNC and LNV processes, after applying the basic cuts, b-jet vetoes and the first BDT cut, are presented in Figs.~\ref{fig:obsLNCvsLNVmN20} and \ref{fig:obsLNCvsLNVmN50} for the $m_N$ = 20 GeV and  50 GeV benchmarks, respectively. 

\begin{figure}[h]
\includegraphics[scale=0.09]{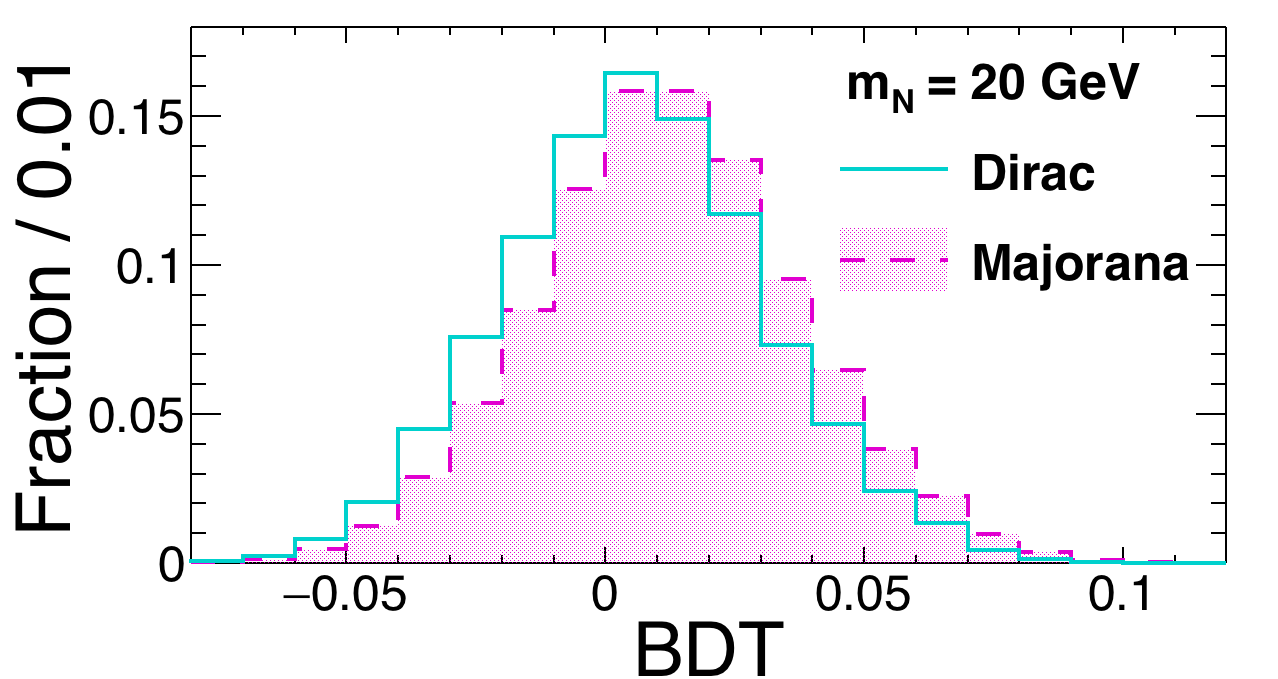}
\includegraphics[scale=0.09]{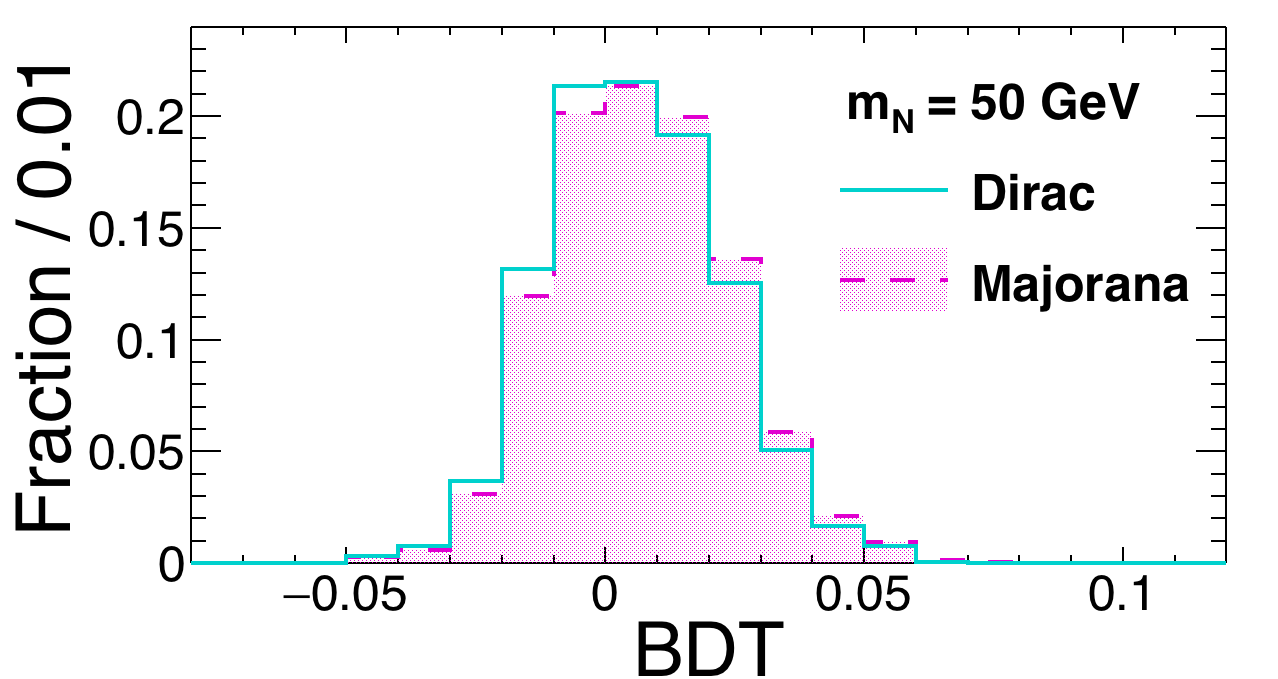}
\caption{
Distributions of the BDT response in the second BDT analysis, for the Dirac hypothesis (dashed line with filled area) and the Majorana scenario (solid line), for the  benchmarks $m_N$ = 20 GeV (left) and 50 GeV (right).
}
\label{fig:BDTmajVSdrc}
\end{figure}

In the second BDT analysis, the simulated data for the Majorana scenario (LNV+LNC+SM background after the first BDT cut) are input as the signal sample, while the simulated data for the Dirac hypothesis (LNC+SM background, with matching $s_D$) are input the as the background sample in the TMVA training and testing processes. The BDT distribution will then indicate the differences between the Majorana scenario and the Dirac hypothesis. Fig.~\ref{fig:BDTmajVSdrc} shows the BDT response for the Majorana case and the Dirac hypothesis for $m_N$ = 20 GeV (left) and 50 GeV (right) benchmark points. Comparing the left and right plots, we see that the 
histograms are more separated in the left plot, leading to a better BDT cut efficiency and thus a higher significance for the $m_N$ = 20 GeV benchmark point.

With an optimized second BDT cut of about 0.020, the Majorana case ends up with 46.1 events, while the Dirac hypothesis has 34.1 events. After defining the number of events corresponding to the excess in the Majorana case from the Dirac hypothesis as the ``signal'' events $N_s$, and the number of events corresponding to the Dirac hypothesis as the ``background" events $N_b$, 
the statistical significance for distinguishing the Majorana scenario from the Dirac hypothesis can be calculated as $\mathit{SS} = N_s/\sqrt{N_s+N_b} = (46.1 - 34.1)/\sqrt{46.1} \approx 1.8$.

\begin{figure}[h]
\includegraphics[scale=0.09]{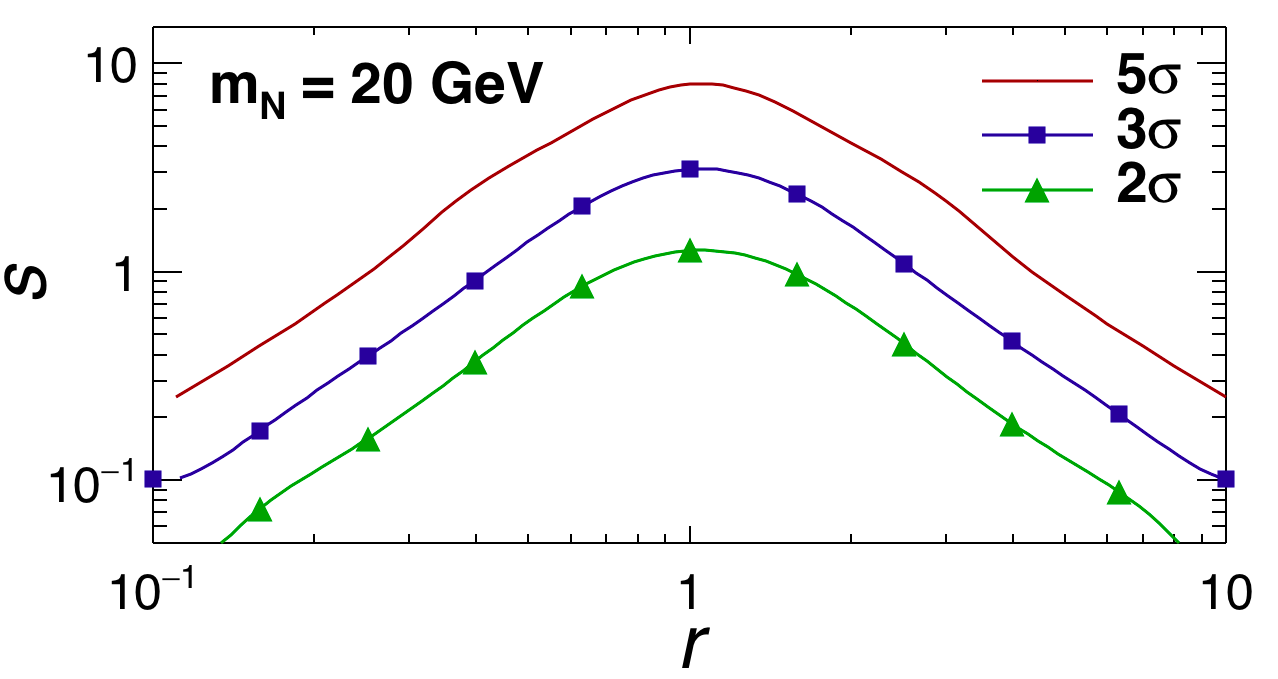}
\includegraphics[scale=0.09]{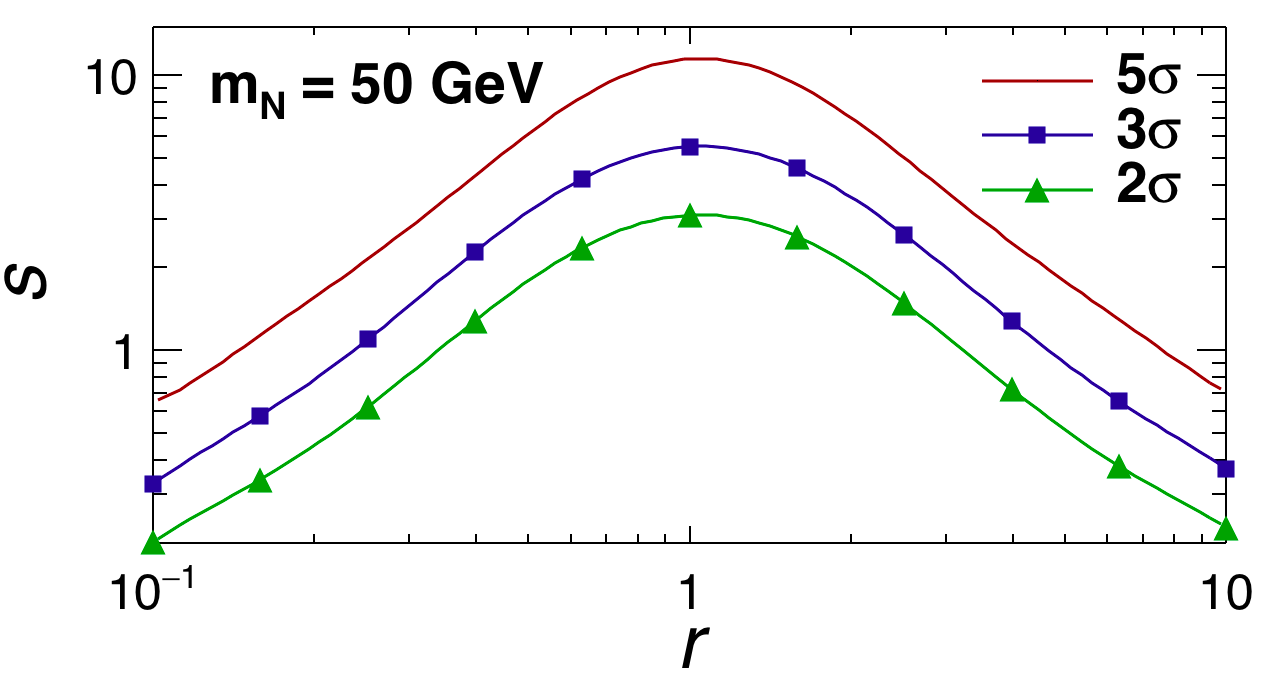}
\caption{Confidence levels of distinguishing between Dirac and Majorana neutrinos for $m_N$ = 20 (left) and 50 (right) GeV. }
\label{fig:sgfDrcVSmaj}
\end{figure}

This three-step method can be extended to the cases where $r \neq 1 $. 
For a given value of the parameter $s$, when $r \gg1$ or $r\ll 1$, from Table~\ref{tab:Nsig} one can see that  the relative number of events for different trilepton states will be quite different in the Majorana scenario, but not so in the Dirac scenario. This feature helps in the Majorana/Dirac discrimination and 
results in 
higher significances.
Fig.~\ref{fig:sgfDrcVSmaj} shows the confidence levels for distinguishing between Majorana and Dirac scenarios, obtained with the above three-step method. When $r \approx 1$, one can have significances near $ 5.0\,\sigma\,  (3.0\,\sigma)$ when $s \geq 7.93\, (3.10)$ in the $m_N = 20$ GeV benchmark, and when $s \geq 11.44\, (5.47)$ in the $m_N = 50$ GeV benchmark. As $r \approx$ 10, to reach the same significance, the parameter $s$ can be as low as 0.25 (0.10) for $m_N = 20$ GeV, and as low as 0.72 (0.38) for $m_N = 50$ GeV.


\section{Conclusions}
\label{sec:Summary}

We present a method to detect and distinguish Dirac and Majorana heavy sterile neutrinos with masses near or below the $W$ boson mass, 
based on the experimental search of the purely leptonic decays $W^\pm \to e^\pm  e^\pm  \mu^\mp \nu$ and $W^\pm\to \mu^\pm  \mu^\pm  e^\mp \nu$  at the 14 TeV LHC, which are induced by a heavy neutrino in the intermediate state. The method is based on both a Cut-and-Count (CC) as well as a Multi-Variate Analysis (MVA). 
Our analysis sets discovery limits on the heavy-to light lepton mixings $ |U_{Ne}|^2$ and $|U_{N\mu}|^2$, which we  express here in terms of the parameters $s = 2\times10^6\times\, |U_{Ne}^{}U_{N\mu}^{}|^2/ \left ( |U_{Ne}|^2+|U_{N\mu}|^2 \right)$ and $r = |U_{Ne}/U_{N\mu}|^2$. The discovery potential of heavy Dirac neutrinos depends on $s$ only, while in the case of Majorana neutrinos it depends on both $s$ and $r$.
The best results are found with the MVA method; nevertheless,  the performance of the 
MVA method becomes close to that of the CC method for small $s$ values, due to the smaller number of signal events.
We use two benchmark points for the heavy neutrino mass: $m_N = 20$ GeV and 50 GeV, and assume an LHC integrated luminosity of $3000~\text{fb}^{-1}$.

Using the MVA method, we find that Dirac sterile neutrinos can be discovered with a significance of at least $ 5.0\,\sigma\,  (3.0\,\sigma)$ when $s \geq 0.55\, (0.25)$  in the case \hbox{$m_N = $ 20 GeV}, or when $s \geq 1.02\, (0.55) $ in the case \hbox{$m_N = $ 50 GeV}. Let us recall that, for $r=1$, the mixings are
$|U_{Ne}|^2 = |U_{N\mu}|^2 = s\times 10^{-6}$.

For Majorana sterile neutrinos, if $r= 1$, the same level of significance can be reached for lower values of $s$ because now the events come from both the LNC and LNV processes. Indeed, a significance of $ 5.0\,\sigma\,  (3.0\,\sigma)$ is 
reached when $s \geq 0.24\, (0.11)$  in the case $m_N = $ 20 GeV, or when $s \geq 0.46\, (0.25) $ in the case $m_N = $ 50 GeV. For the same $s$ but $r \neq 1$ the significances also become larger for a Majorana neutrino, due to the larger number of events:  when \emph{e.g.} $r = 10$,  a significance of  $ 5.0\,\sigma\,  (3.0\,\sigma)$ is reached when $s \geq 0.08\, (0.03) $ for the benchmark $m_N = $ 20 GeV, and $s \geq 0.16\, (0.09) $ for the benchmark $m_N = $ 50 GeV. Let us now recall that, for $r=10$, the mixings are
$|U_{Ne}|^2 = 10\, |U_{N\mu}|^2 = 5.5\, s\times 10^{-6}$.

Finally, the Dirac and Majorana hypotheses can be distinguished from each other at $ 5.0\,\sigma\,  (3.0\,\sigma)$ level of significance when $r \approx 1$ provided that  $s \geq 7.93\, (3.10)$ for the benchmark $m_N = 20$ GeV, and $s \geq 11.44\, (5.47)$ for the benchmark $m_N = 50$ GeV.  For $r\ll 1$ or $r\gg 1$, lower values of $s$ are required: as $r \approx$ 10, to reach the same significance the parameter $s$ can be as low as 0.25 (0.10) for $m_N = 20$ GeV, or as low as  0.72 (0.38) for $m_N = 50$ GeV.

\begin{acknowledgments}
We want to thank Jue Zhang for valuable help. K.W. was supported by the International Postdoctoral Exchange Fellowship Program (No.90 Document of OCPC, 2015);  C.S.K. by the NRF grant funded by the Korean government of the MEST (No. 2016R1D1A1A02936965); 
and C.D. by Chile grants Fondecyt No.~1130617, 1170171, Conicyt ACT 1406 and PIA/Basal FB0821.
\end{acknowledgments}

\appendix

\section{Validation for Fake Lepton Simulation}
\label{sec:FL}

\begin{figure}[h]
\centering
\includegraphics[scale=0.4]{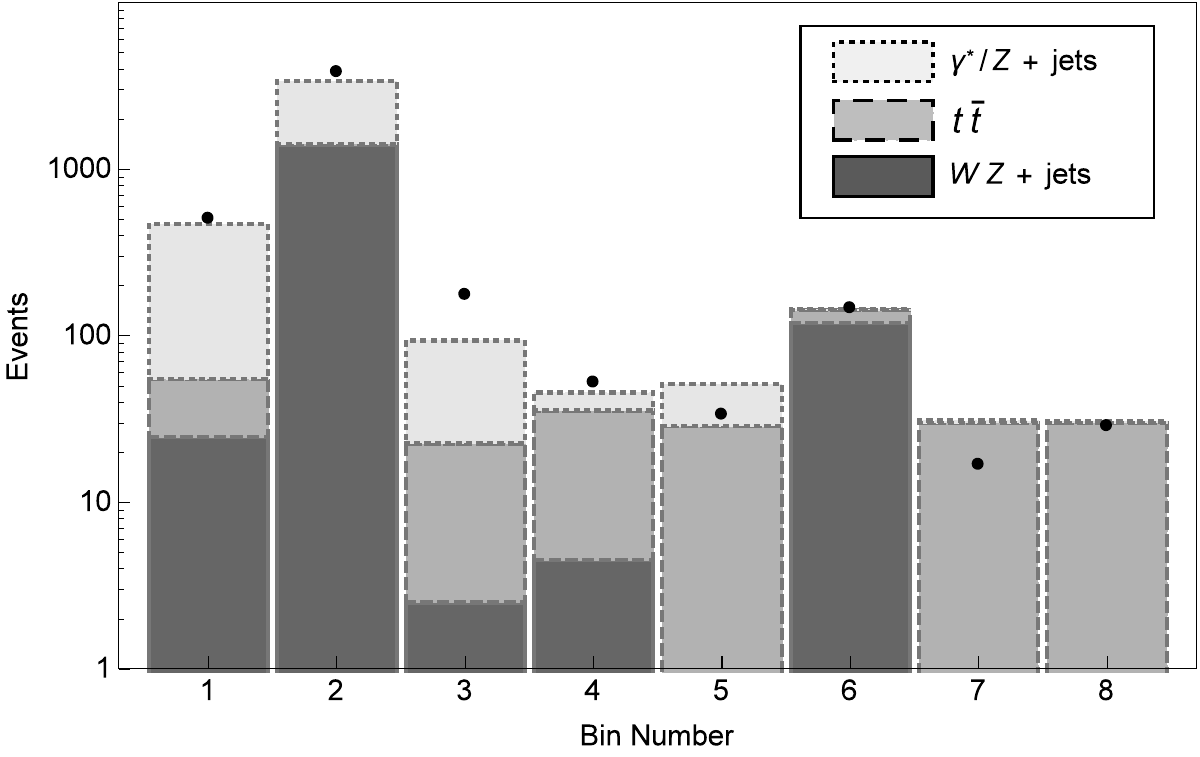}
\caption{Validation results for fake lepton simulation. Black dots indicate experimental results in Ref. \cite{Chatrchyan:2014aea}. Our simulated results for $\gamma^*_{}/Z$+jets, $t\bar{t}$ and $WZ$+jets are given by up light gray bars, middle brown bars and bottom pink bars, respectively. Eight bin categories are: (1) 0-bjet, 1-OSSF, $M_{\ell^+_{}, \ell^-_{}}^{} < 75 ~{\rm GeV}$; (2) 0-bjet, 1-OSSF, $|M_{\ell^+_{}, \ell^-_{}}^{} - M_Z^{} | < 15 ~{\rm GeV}$; (3) 0-bjet, 1-OSSF, $M_{\ell^+_{}, \ell^-_{}}^{} >105 ~{\rm GeV}$; (4) 0-bjet, 0-OSSF; (5-8) are the same as the first four bins, but with at least one b-jet.}
\label{fg:validation}
\end{figure}

In this appendix, we present our validation results for the fake lepton simulation used in this work. We follow closely the same validation done in Ref.~\cite{Izaguirre:2015pga}, and find out that using their modeling parameters, the simulation results can indeed be consistent with the experimental results given in Ref.~\cite{Chatrchyan:2014aea}. Specifically, we take $r_{10}^{} = 1$, $\mu = 0.5$, $\sigma = 0.3$ and $\epsilon_{200}^{} = 4.6 \times 10^{-3}_{}$. In fact, the suggested mistag rate of $\epsilon_{200}^{} = 4.6 \times 10^{-3}_{}$ coincides with the ``rule-of-thumb" introduced in Ref.~\cite{Sullivan:2010jk}, i.e., isolated electrons and muons from heavy-flavor decay are about $1/200$ times the rates of $b$ and $c$ quark production. For the other input parameters of $r_{10}^{}$, $\mu$ and $\sigma$, the authors of Ref. \cite{Izaguirre:2015pga} find that varying them does not substantially change the fitting to the data, provided the overall fake efficiency of $\epsilon_{200}^{}$ remains fixed.

Our validation results are shown in Figure \ref{fg:validation}. Each bin represents an event category according to the CMS trilepton search given in Ref.~\cite{Izaguirre:2015pga}, namely, (1) 0-bjet, 1-OSSF, $M_{\ell^+_{}, \ell^-_{}}^{} < 75 ~{\rm GeV}$; (2) 0-bjet, 1-OSSF, $|M_{\ell^+_{}, \ell^-_{}}^{} - M_Z^{} | < 15 ~{\rm GeV}$; (3) 0-bjet, 1-OSSF, $M_{\ell^+_{}, \ell^-_{}}^{} >105 ~{\rm GeV}$; (4) 0-bjet, 0-OSSF; (5-8) are the same as the first four bins, but with at least one b-jet. The actual experiment results are indicated by black dots, while our simulated results are given by upper light bars, middle dark bars and bottom light bars for the processes $\gamma^*_{}/Z$+jets, $t\bar{t}$ and $WZ$+jets, respectively. As one can see, our results agree with the experimental results reasonably well within the statistical uncertainties, especially in bin-4, whose selection criteria mostly resemble the ones stated in our main text. Moreover, a good agreement with the results given in Fig.~10 of Ref.~\cite{Izaguirre:2015pga} is also found, although in some bins we differ in the individual fractions of events from different processes.



\end{document}